\documentclass[11pt,a4paper,english,twoside]{article}

\usepackage{a4wide}
\usepackage{amssymb, amsmath}
\usepackage{graphicx}
\usepackage[all]{xy}
\usepackage{hyperref}
\hypersetup{linktocpage}
\usepackage{enumerate}
\usepackage{xcolor}
\usepackage{dsfont}
\usepackage{empheq}
\usepackage{cite}
\usepackage{float}

\usepackage{soul}

\usepackage{enumitem}
\usepackage{hhline}

\newcommand{\beq}{\begin{equation}}
\newcommand{\eeq}{\end{equation}}
\def\bea#1\eea{\begin{align}#1\end{align}}
\def\beal#1\eeal{\begin{subequations}\begin{align}#1\end{align}\end{subequations}}
\newcommand{\nn}{\nonumber}
\newcommand{\w}{\wedge}

\newcommand{\R}{\mathcal{R}}

\renewcommand{\i}{\ensuremath{\textnormal{i}}}

\def\del {\partial}
\def\d {{\rm d}}
\def\mmm {\mathcal{M}}
\def\tV {\tilde{V}}
\def\hta {\hat{\tau}}
\def\hr {\hat{\rho}}

\begin{document}
\numberwithin{equation}{section}

\begin{titlepage}

\begin{center}

\phantom{DRAFT}

\vspace{2.4cm}

{\LARGE \bf{The web of swampland conjectures \vspace{0.4cm}\\ and the TCC bound}}\\

\vspace{2.2 cm} {\Large David Andriot$^{1}$, Niccol{\`o} Cribiori$^{1}$, David Erkinger$^{2}$}\\
 \vspace{0.9 cm} {\small\slshape $^1$ Institute for Theoretical Physics, TU Wien\\
Wiedner Hauptstrasse 8-10/136, A-1040 Vienna, Austria}\\
 \vspace{0.2 cm} {\small\slshape $^2$ Mathematical Physics Group, University of Vienna\\
Boltzmanngasse 5, 1090 Vienna, Austria}\\
\vspace{0.5cm} {\upshape\ttfamily david.andriot@tuwien.ac.at; niccolo.cribiori@tuwien.ac.at; david.josef.erkinger@univie.ac.at}\\

\vspace{2.8cm}

{\bf Abstract}
\vspace{0.1cm}
\end{center}

\begin{quotation}
\noindent We consider the swampland distance and de Sitter conjectures, of respective order one parameters $\lambda$ and $c$. Inspired by the recent Trans-Planckian Censorship conjecture (TCC), we propose a generalization of the distance conjecture, which bounds $\lambda$ to be a half of the TCC bound for $c$, i.e.~$\lambda \geq \frac{1}{2}\sqrt{\frac{2}{3}}$ in 4d. In addition, we propose a correspondence between the two conjectures, relating the tower mass $m$ on the one side to the scalar potential $V$ on the other side schematically as $m\sim |V|^{\frac{1}{2}}$, in the large distance limit. These proposals suggest a generalization of the scalar weak gravity conjecture, and are supported by a variety of examples. The lower bound on $\lambda$ is verified explicitly in many cases in the literature. The TCC bound on $c$ is checked as well on ten different no-go theorems, which are worked-out in detail, and $V$ is analysed in the asymptotic limit. In particular, new results on 4d scalar potentials from type II compactifications are obtained.
\end{quotation}

\end{titlepage}

\newpage

\tableofcontents

\section{Introduction and summary}

The swampland program aims at characterising low energy effective theories of quantum gravity. In particular, it provides general criteria for models to descend from string theory, and be part of the landscape, or not and be part of the swampland. The program allows in principle to discriminate among various phenomenological models, which otherwise would appear on the same footing, and is thus of interest to phenomenology, especially for cosmological or particle physics models. However, most of the existing criteria are for now at the state of conjectures. Still, these conjectures are heavily discussed, tested and refined. A recent review can be found in \cite{Palti:2019pca}. Beyond a mere list of criteria, there is a hope to understand deeper structures in quantum gravity and effective theories, that would be responsible for all of these characterisations. As a first step in this direction, it appears that most of these conjectures are related to each other, forming more a web of conjectures rather than a list. In this paper, we are interested in two conjectures and their refinements, the de Sitter conjecture and the distance conjecture, as well as in tentative relations between them.\\

The first conjecture analysed here is the swampland de Sitter conjecture, originally proposed in \cite{Obied:2018sgi}. It is notoriously difficult to get de Sitter solutions in string theory, meaning backgrounds with a de Sitter space-time, in a well-controlled manner. This is particularly known in the classical regime, where many no-go theorems were established against such solutions. We give a brief overview of this topic in section \ref{sec:sumc}. This situation has motivated the proposal of a swampland conjecture. It applies to a $d$-dimensional theory, $d>2$, of scalar fields $\varphi^i$ coupled to gravity, of the form
\beq
{\cal S}=  \int \d^d x \sqrt{|g_{d}|} \left(\frac{M_p^2}{2} {\cal R}_{d} - \frac{1}{2} g_{ij} \del_{\mu} \varphi^i \del^{\mu} \varphi^j - V(\varphi) \right) \ , \label{Sddintro}
\eeq
with $M_p$ the Planck mass (set to 1 in the following), $\mu$ the $d$-dimensional indices, $g_{ij}$ the field space metric and $V(\varphi)$ the scalar potential. For such a theory to be a low energy effective theory of string theory, the de Sitter conjecture \cite{Obied:2018sgi} states that its potential should obey the following inequality (in Planckian units)
\beq
|\nabla V| \geq c\, V \ , \quad c \sim \mathcal{O}(1) \ , \label{V'/V}
\eeq
where $|\nabla V|= \sqrt{g^{ij}\del_i V \del_j V}$. At an extremum of the potential where $\nabla V=0$, from \eqref{Sddintro} one obtains a static solution verifying ${\cal R}_{d} =V \times 2d/(d-2)$. The inequality \eqref{V'/V} then implies the absence of any such de Sitter solution. Beyond the extrema, this inequality is about the slope of the potential. This has important consequences for cosmological models, in particular for models of inflation in the early universe, as well as for the description of the late universe, for which quintessence is advocated by this conjecture. These implications were first explored in \cite{Agrawal:2018own} and then much debated. Further arguments led to the idea that this conjecture was too strong and needed a refinement, to allow in particular for de Sitter maxima. Such a refinement was proposed in \cite{Ooguri:2018wrx} (see also \cite{Andriot:2018wzk, Garg:2018reu, Andriot:2018mav, Rudelius:2019cfh}, and the recent \cite{Luben:2020wim}). Additionally, the paper \cite{Ooguri:2018wrx} brought the important idea that the conjecture \eqref{V'/V} would only be valid in the asymptotics of field space, generalising the result of \cite{Dine:1985he}. Considering such large field distances corresponds to having specific regimes of string theory, for instance a classical, perturbative regime, which is the one where we can restrict to 10d supergravities, as we will do in section \ref{sec:dSsection}. This idea of considering the large field limit fits nicely with the fact that many no-go theorems were obtained precisely in this framework.

More generally, it is fair to say that the initial de Sitter conjecture had two drawbacks: it was lacking of a deeper physical motivation and it did not provide a definite value for $c$, which is crucial for cosmology. Interestingly, the recent Trans-Planckian Censorship conjecture (TCC) proposed in \cite{Bedroya:2019snp} answers these two points (see also \cite{Saito:2019tkc}). Based on some physical arguments regarding quantum fluctuations, an inequality similar to \eqref{V'/V}, with a specific value for the bound parameter, was derived in an asymptotic or large field distance limit (assuming a positive, rolling potential). Denoting the field space geodesic distance from a point $P$ to $Q$ as ${\cal D}(P,Q)$, or simply ${\cal D}=|\varphi - \varphi_i|$ for a canonically normalised field $\varphi$, the derivation of \cite{Bedroya:2019snp} led to
\beq
0 < V(\varphi) < V_0\, e^{- c_0\, {\cal D}}  \quad \Rightarrow \quad \left\langle \frac{|V'|}{V} \right\rangle_{{\cal D} \rightarrow \infty}\, \geq\, c_0\, = \, \frac{2}{\sqrt{(d-1)(d-2)}} \ , \label{TCC}
\eeq
with constants $V_0$, $\varphi_i$, and where one averages the ratio $|V'|/V$, considered here for a single field, and takes the limit of large distances. We reproduce and generalize this derivation around \eqref{distconjus2}. This proposal was also extended to the multi-field case \cite{Bedroya:2019snp}. The inequality \eqref{TCC} thus provides a lower bound value for $c$ in \eqref{V'/V}, in particular
\beq
\mbox{TCC bound:}\qquad c \geq \sqrt{\frac{2}{3}} \approx 0.8165 \qquad {\rm in}\ d=4 \ .\label{TCCbound}
\eeq
This value was verified in \cite{Bedroya:2019snp} to hold for various examples coming from, or related to, string theory, and in particular for three no-go theorems on classical de Sitter solutions. The TCC had further consequences, in particular an upper bound on the lifetime of an expanding universe, that was shown to be a half of the scrambling time in \cite{Aalsma:2020aib}; we will not consider here these stability-related aspects. Finally, the cosmological implications of the TCC, especially for inflation, were heavily discussed, and we refrain here from commenting on those. Regarding the actual number in \eqref{TCCbound}, we simply note that the corresponding single field exponential quintessence model would be allowed by observational data in a $3\sigma$ region, as displayed in figure 2 of \cite{Akrami:2018ylq} (see also the recent \cite{Heisenberg:2020ywd}). We will come back to this bound in detail.\\

The second conjecture of interest is the (refined) swampland distance conjecture \cite{Ooguri:2006in, Klaewer:2016kiy, Baume:2016psm}. This conjecture first asserts that at any point in the scalar moduli space (or a more general field space \cite{Lust:2019zwm}) of a string compactification, one can find another point at an arbitrarily large geodesic distance ${\cal D}$. In addition, in the asymptotic limit ${\cal D} \rightarrow \infty$, the conjecture claims that an infinite tower of states appears and becomes massless, with a mass-scale $m$ expressed in Planckian units as
\beq
m(\varphi) \approx m_i\, e^{-\lambda\, {\cal D}} \ \text{for}\ {\cal D} \rightarrow \infty\ ,\quad \lambda \sim \mathcal{O}(1) \ ,\label{distconj}
\eeq
with the initial mass-scale $m_i=m(\varphi_i)$. The refined conjecture states that the tower appears for a super-Planckian distance ${\cal D}$, setting $\lambda \sim \mathcal{O}(1)$. These new light states signal a breakdown of the effective field theory description. As we will report in section \ref{sec:dist}, this conjecture has been successfully checked in various works, determining the value of $\lambda$ in different examples.

Interestingly, both conjectures display a number of order one, $c$ and $\lambda$. On general grounds, if conjectures are related, then the various numbers entering them should be related as well. As we will review in section \ref{sec:weblit}, relations between the distance conjecture and the de Sitter conjecture have been proposed, starting with \cite{Ooguri:2018wrx}. We would infer a relation between $c$ and $\lambda$. A natural question is then whether the TCC bound \eqref{TCCbound} gets translated into a lower bound on $\lambda$. These are the topics of this work, and we now summarize our results.

\subsubsection*{Summary of the results}

In section \ref{sec:dSsection}, we focus on no-go theorems against classical de Sitter solutions and their related $c$ value. The aim is to check to larger extent, namely ten no-go theorems, whether the TCC bound \eqref{TCCbound} holds. To this end, we first need to translate known no-go theorems into a condition on a four-dimensional potential, and furthermore such a condition should be valid away from an extremum. We present the necessary tools in section \ref{sec:framework} and in appendix \ref{ap:4formgeneral}. We use a previously known potential $V(\rho,\tau,\sigma)$ \cite{Andriot:2018ept}, but provide for it a proper treatment of the 4d four-form fluxes, modifying the $F_5$ and $F_6$ terms. We also derive a new scalar potential $V(r,\tau)$ depending on a radius and a dilaton scalar fields. For each no-go theorem, we then give in section \ref{sec:nogolist} a 4d formulation in terms of an inequality of the form
\beq
a\, V + \sum_{i=1}^3 b_i\, \del_{\hat{\varphi}^i} V \leq 0 \ ,\quad a>0\ ,\label{combinogointro}
\eeq
from which we deduce a value for $c$. The approach and results are summarized and commented in section \ref{sec:sumc}, in particular in Table \ref{tab:c}. A net result is that the TCC bound \eqref{TCCbound} is always satisfied, with some saturation cases. While this is expected from the swampland perspective described above, it remains a surprisingly successful check, since we only considered purely classical supergravity potentials, without considering any quantum gravity argument, nor any averaging or limit.

In section \ref{sec:dist}, we report on the literature where the distance conjecture \eqref{distconj} was verified, listing in particular in Table \ref{tab:lambda} the values of $\lambda$ obtained so far and computing a few more. We also harmonise various conventions on the field distance in the literature, compatible with the kinetic terms of \eqref{Sddintro}, to be able to compare such $\lambda$ values. To our surprise, we note that all these examples (of compactifications to 4d) verify the following bound
\beq
\lambda\ \geq\ \frac{1}{2}\, \sqrt{\frac{2}{3}}\ ,\label{lambdaboundintro}
\eeq
i.e.~a half of the TCC bound \eqref{TCCbound}, with several cases of saturation. We additionally note that other values in Table \ref{tab:c} and \ref{tab:lambda} match up to the same $\frac{1}{2}$ factor. We stressed already that the compactification setups on both sides, as well as the scalar fields involved, are very different. The approaches to get these numbers are also completely different, there is thus at first sight no obvious reason for such a correspondence, except from a swampland perspective.

Indeed, we then discuss in section \ref{sec:web} the relations between conjectures. After reviewing in section \ref{sec:weblit} ideas and further conjectures given in the literature, we make some inspired proposals in section \ref{sec:prop}, based on the data collected in the previous sections. We first propose in section \ref{sec:propdist} to generalize the distance conjecture to a form analogous to \eqref{TCC}, namely
\beq
 0 < m \leq m_0\ e^{- \lambda_0 \, {\cal D}} \quad \text{for} \ {\cal D} \rightarrow \infty \ , \qquad \text{with}\ \lambda_0 = \frac{1}{2}\, c_0\, = \, \frac{1}{\sqrt{(d-1)(d-2)}} \ .\label{distconjusintro}
\eeq
We further prove the implication (without assumption on the sign of $m'$)
\beq
\eqref{distconjusintro} \ \Rightarrow\ \left\langle \frac{|m'|}{m} \right\rangle_{{\cal D} \rightarrow \infty}\, \geq\, \lambda_0 \ . \label{distconjus2intro}
\eeq
When specified to the particular case of the standard distance conjecture \eqref{distconj}, the bound \eqref{distconjus2intro} reproduces the observed bound \eqref{lambdaboundintro} on $\lambda$, namely
\beq
m = m_i\ e^{- \lambda \, {\cal D}} \quad \Rightarrow \quad \lambda\, \geq\, \lambda_0 \ . \label{stand}
\eeq
To avoid confusion, let us mention that another bound on $\lambda$ was briefly discussed in section 6 of \cite{Bedroya:2019snp}: we never consider the latter in this paper, our bound $\lambda_0$ given in \eqref{distconjusintro} is different. Finally, the proposal \eqref{distconjusintro} interestingly allows more general functions $m$ than in \eqref{stand}, e.g.~some including logarithmic corrections. A more detailed presentation can be found in section \ref{sec:propdist}.

In section \ref{sec:propswgc}, we then propose the inequality in the right-hand side of \eqref{distconjus2intro} as a natural generalization of the scalar weak gravity conjecture \cite{Palti:2017elp}. In the case where no average or limit is needed, it boils down to
\beq
(\del_{\varphi} m)^2 \geq \lambda_0^2\ m^2 \ .
\eeq
A further proposal generalizing that of \cite{Gonzalo:2019gjp}, and a discussion, are given in section \ref{sec:propswgc}.

Finally, we propose in section \ref{sec:proprel} a relation between the generalized distance conjecture \eqref{distconjusintro} and the de Sitter conjecture in TCC form \eqref{TCC}. It takes the form of a map between two compactification setups, together with two field space directions, in each of which one conjecture applies. The map relates the mass $m$ on one side to the potential $V$ on the other side in a large field distance limit, as follows
\beq
\frac{m}{m_i} \simeq \left|\frac{V}{V_i}\right|^{\frac{1}{2}} \quad \mbox{for}\ {\cal D} \rightarrow \infty \ , \label{ourconjintro}
\eeq
up to constants $m_i, V_i$. In the case where $m$ and $V$ are exponentials along the selected field directions, in the large field limit, one deduces the relation
\beq
\lambda = \frac{1}{2} \, c\ \geq\ \lambda_0 \ ,\label{relclambdaintro}
\eeq
as further discussed in section \ref{sec:checkmap}. The relation \eqref{relclambdaintro} can be observed by comparing examples of Table \ref{tab:c} and \ref{tab:lambda}. In addition, the map \eqref{ourconjintro} includes as a particular case (a de Sitter extension of) the strong anti-de Sitter distance conjecture of \cite{Lust:2019zwm}. A more detailed discussion can be found in sections \ref{sec:proprel} and \ref{sec:checkmap}.

The two conjectures considered in this paper show an important dependence on a selected field direction in field space. On the distance conjecture side, the value of $\lambda$ and the nature of the tower of states are dependent on this direction. On the de Sitter conjecture side, we note a dependence on a specific field direction $\varphi_2$, given by the linear combination of fields entering the no-go inequality \eqref{combinogointro}. This inequality can be rewritten as follows
\beq
a\, V + \sum_{i} b_i\, \del_{\hat{\varphi}^i} V \leq 0 \quad \leftrightarrow\quad  c\, V + \del_{\varphi_2} V \leq 0 \ .
\eeq
We verify in several examples that the potential then takes the following asymptotic form
\beq
V = V_i\ e^{- c \, {\cal D}} \quad \mbox{for}\ {\cal D} \rightarrow \infty \ ,
\eeq
where ${\cal D}=|\varphi_2 - \varphi_i|$. This precise exponential form is a non-trivial check of the proposed map \eqref{ourconjintro}, to exponential masses $m$. It also gives a new understanding of the number $c$, verifying the relation \eqref{relclambdaintro}. We detail these ideas and checks in section \ref{sec:checkmap}.

The examples of Table \ref{tab:c} and \ref{tab:lambda} depend on the compactification setups and the specific field directions just mentioned. These aspects can be very different, but pairs of such examples can still be related by the proposed map, in large field limits. This is reminiscent of a duality, and we further comment on this idea at several places in section \ref{sec:web}.

\section{No-go theorems on classical de Sitter and $c$ values}\label{sec:dSsection}

De Sitter string backgrounds are backgrounds of string theory with a de Sitter space-time. Here, we specialize to a 10d space-time split as 4d de Sitter times a 6d compact manifold. Classical string backgrounds are those corresponding to a classical regime of string theory. In practice, these are solutions of a 10d supergravity theory, for which one verifies that $\alpha'$ corrections can be neglected (low energy) and that the string coupling $g_s$ is small (weak coupling). We will mainly consider de Sitter solutions of 10d type II supergravities with $D_p$-branes and orientifold $O_p$-planes, and will also briefly mention heterotic string at order $({\alpha'})^0$. It turns out that it is difficult to obtain such classical de Sitter solutions. Indeed, several no-go theorems against them have been derived, under specific assumptions. In this section, we recast known no-go theorems on classical de Sitter string backgrounds in the form of an inequality of the type \eqref{V'/V} and we compute the corresponding value of $c$. We develop the necessary tools and discuss subtleties in the process. In \ref{sec:sumc} we briefly review the topic, giving a list of the known no-go theorem we will be looking at, and we present the problem to be addressed. We then summarize our results, namely the values of $c$, and comment on them. In \ref{sec:framework} we present the specific framework we will be working in and introduce all necessary ingredients. In \ref{sec:nogolist} we reformulate the no-go theorems in a 4d language, obtaining for each of them an inequality of the type \eqref{V'/V} and computing the associated value of $c$.

\subsection{Summary and $c$ values}\label{sec:sumc}

It is notoriously difficult to find a classical de Sitter background of string theory. The typical strategy consists in finding a de Sitter solution in a 10d supergravity theory and verifying then whether or not such a solution is in the appropriate low energy and weak coupling regime, with respect to string theory. The first step is already non-trivial and many no-go theorems or constraints have been established against de Sitter solutions. One can see \cite{Danielsson:2018ztv, Andriot:2019wrs} or the introduction of \cite{Andriot:2018ept} for recent reviews and references. Most of the focus has been in type II supergravities with $D_p$-branes and $O_p$-planes sources. It is important to distinguish whether the sources are all parallel, or whether they are intersecting, i.e.~wrapping different directions. Indeed, few de Sitter solutions of 10d type IIA/B supergravities have been found with intersecting sources \cite{Caviezel:2008tf, Flauger:2008ad, Danielsson:2009ff, Caviezel:2009tu, Danielsson:2010bc, Danielsson:2011au, Roupec:2018mbn}, but none with parallel ones.\footnote{Solutions with parallel sources might be found when allowing for different source boundary conditions \cite{Cordova:2018dbb, Cribiori:2019clo, Cordova:2019cvf}, that we do not consider here.} Accordingly, one verifies that some constraints get much stronger with parallel sources compared to intersecting ones \cite{Andriot:2016xvq, Andriot:2017jhf}. In Table \ref{tab:solnogo}, we summarize most of the constraints known on the existence of solutions with parallel sources, on which we focus in this paper. The corresponding table for the intersecting case can be found in \cite{Andriot:2019wrs}. It excludes actually the same cases, but it leaves more freedom in the required ingredients. Once one has few supergravity de Sitter solutions, one may wonder about their validity as classical string backgrounds. Several works argued against it \cite{Roupec:2018mbn, Junghans:2018gdb, Banlaki:2018ayh, Andriot:2019wrs, Grimm:2019ixq} in specific settings suggesting for now that no classical de Sitter string background exists, in agreement with the swampland inequalities \eqref{V'/V} and \eqref{TCC} of \cite{Ooguri:2018wrx, Bedroya:2019snp}, but loopholes were also indicated in \cite{Hebecker:2018vxz, Junghans:2018gdb, Andriot:2019wrs} (see also \cite{Andriot:2020wpp}).
\begin{table}[H]
\begin{center}
\begin{tabular}{|c||c|c|}
    \hline
     & \multicolumn{2}{|c|}{} \\[-8pt]
     & \multicolumn{2}{|c|}{A de Sitter solution requires $T_{10}>0$ (\ref{nogo1}) and}\\[3pt]
    \hhline{~||~~} %\cdashline{2-3}
    $p$ &  ${\cal R}_6 \geq 0$ & ${\cal R}_6 <0$ \\[3pt]
    \hhline{===}
     & \multicolumn{2}{|c|}{} \\[-8pt]
    3 & \multicolumn{2}{|c|}{(\ref{nogo4})} \\[3pt]
    \hhline{-||--}
     & & \\[-8pt]
    4 & $\phantom{f^{||}{}_{\bot\bot}\, (\ref{nogo5}), (\ref{nogo6}), (\ref{nogo9}),\, f^{\bot}{}_{\bot ||}\, (\ref{nogo7}), (\ref{nogo8}),}$ & $F_{6-p}\, (\ref{nogo2}),$ \\[3pt]
    \hhline{-||~~}
     & & \\[-8pt]
    5 & (\ref{nogo3}) & $f^{||}{}_{\bot\bot}\, (\ref{nogo5}), (\ref{nogo6}), (\ref{nogo9}),\, f^{\bot}{}_{\bot ||}\, (\ref{nogo7}), (\ref{nogo8}),$ \\[3pt]
    \hhline{-||~~}
     & & \\[-8pt]
    6 &  & $\mbox{linear combi}\, (\ref{nogo5}), (\ref{nogo6})$ \\[3pt]
    \hhline{-||--}
     & & \\[-8pt]
    7 &  &  \\[3pt]
    \hhline{-||~~}
     & & \\[-8pt]
    8 & (\ref{nogo2}), (\ref{nogo3}) & (\ref{nogo2}) \\[3pt]
    \hhline{-||~~}
     & & \\[-8pt]
    9 &  &  \\[3pt]
    \hline
\end{tabular} \caption{Most of the constraints known on the existence of type II supergravity de Sitter solutions with parallel $D_p/O_p$ sources (with single size $p$). Few more constraints, and an analogous table for intersecting sources, can be found in \cite{Andriot:2019wrs}. An empty cell implies the absence of de Sitter solutions, while an entry (flux, etc., defined in section \ref{sec:framework}) means that it is a necessary ingredient to get a de Sitter solution. The ({\bf number.}) refer to the no-go theorems, discussed in this paper, that prove the absence of solution or the need of some ingredient and constraints on it.}\label{tab:solnogo}
\end{center}
\end{table}

In the present work we want to translate all the no-go theorems or constraints indicated in Table \ref{tab:solnogo} into an inequality of the form \eqref{V'/V} and, from that, read-off the value of the parameter $c$. Getting to the inequality \eqref{V'/V} is not straightforward, for at least two reasons. First, several no-go theorems were only derived in 10d, using in particular equations of motion, and one should then translate them into a 4d potential and its first derivatives: this is not always straightforward. Second, the inequality \eqref{V'/V} is an off-shell one, in the sense that it should be valid at any point in field space and not just at critical points of the potential, i.e.~de Sitter solutions in our case. This will lead to some additional complications as we will see.

In the following, we will deal with the two difficulties just mentioned and work-out systematically the translation of the no-go theorems. For each of them, we then compute the parameter $c$, as explained below. We summarize our findings in Table \ref{tab:c}.
\begin{table}[H]
  \begin{center}
    \begin{tabular}{|c|c|c|c||c|}
    \hline
    No-go & Condition & No field & Use of &   \\
    number & for the no-go & in condition & BI & $c$ \\
    \hhline{=====}
     & & & & \\[-8pt]
    \ref{nogo1} & $T_{10} \leq 0$ & $\checkmark$ & &  $\sqrt{2}$ \\[3pt]
    \hhline{----||-}
     & & & & \\[-8pt]
    \ref{nogo2} & $p=7,8$, or $p=4,5,6$ $\&$ $F_{6-p}=0$ & $\checkmark$ & &  $\sqrt{\frac{2(p-3)^2}{3+(p-4)^2}} \geq \sqrt{\frac{2}{3}}$ \\[7pt]
    \hhline{----||-}
     & & & & \\[-8pt]
    \ref{nogo3} & $\R_6 \geq 0$, $p \geq 4$ & $\checkmark$ & &  $\sqrt{\frac{2(p+3)^2}{3+p^2}} > 1$ \\[7pt]
    \hhline{----||-}
     & & & & \\[-8pt]
    \ref{nogo4} & $p=3$ & $\checkmark$ & $\checkmark$ & $2\sqrt{\frac{2}{3}}$ \\[8pt]
    \hhline{----||-}
     & & & & \\[-8pt]
    \ref{nogo5} & $  {\cal R}_{||} + {\cal R}_{||}^{\bot} + \frac{\sigma^{-12}}{2}  |f^{{}_{||}}{}_{{}_{\bot} {}_{\bot}}|^2 \leq 0 $, $p\geq 4$ & $\times$ & & $\sqrt{\frac{2(p-3)}{p-1} } \geq \sqrt{\frac{2}{3}}$ \\[8pt]
    \hhline{----||-}
     & & & & \\[-8pt]
    \ref{nogo6} & $ - 2   \rho^{2} \sigma^{2(p-6)} ( {\cal R}_{||} + {\cal R}_{||}^{\bot}) +  |H^{(2)}|^2   \leq 0 $ & $\times$ & $\checkmark$ & $2\sqrt{\frac{2}{3}}$ \\[8pt]
    \hhline{----||-}
     & & & & \\[-8pt]
    \ref{nogo7} & $\lambda= - \frac{\delta^{cd}   f^{b_{\bot}}{}_{a_{||}  c_{\bot}} f^{a_{||}}{}_{ b_{\bot} d_{\bot}}}{ \tfrac{1}{2} \delta^{ab} \delta^{cd} \delta_{ef}  f^{e_{||}}{}_{a_{\bot} c_{\bot}} f^{f_{||}}{}_{b_{\bot} d_{\bot}} }  \leq 0$, $p\geq 4$ & $\checkmark$ & $\checkmark$ & $\sqrt{\frac{2}{3}}$ \\[8pt]
    \hhline{----||-}
     & & & & \\[-8pt]
    \ref{nogo8} & $\lambda \geq \sigma^{-6} $, $p\geq 4$ & $\times$ &  & $\sqrt{\frac{2p}{p-2}} > 1$ \\[8pt]
    \hhline{----||-}
     & & & & \\[-8pt]
    \ref{nogo9} & $\exists\, a_{||}$ s.t. $f^{a_{||}}{}_{ij} = 0$ $\, \forall i,j \neq a_{||}$, $p\geq 4$ & $\checkmark$ & & $\sqrt{\frac{2}{3}}$ \\[8pt]
    \hhline{----||-}
     & & & & \\[-8pt]
    \ref{nogo10} & Heterotic at order $({\alpha'})^0$ & $\checkmark$ & & $\sqrt{\frac{2}{3}}$  \\[8pt]
    \hline
       \end{tabular}
     \caption{No-go theorems translated into an inequality of the form \eqref{V'/V} and the corresponding value of $c$. The various properties specified for each no-go theorem are discussed in the text. All no-gos are obtained in type II supergravities except the last one.}\label{tab:c}
  \end{center}
\end{table}
\noindent Each no-go theorem relies on some assumptions or conditions, most of the time previously worked-out on-shell, especially when the no-go was obtained with the use of the 10d equations of motion. The additional complication occurring in this case is that such assumptions or conditions, once established off-shell, sometimes become field dependent, as indicated in Table \ref{tab:c}. We believe that such a field dependence is in general not what one expects from a no-go theorem, whose assumptions should be generic. When considering such no-go theorems off-shell, we propose then to take them with caution or view them as weaker. As discussed further in the text however, for some special cases where part of the condition vanishes, one may loose the field dependence making the no-go more interesting. Another point that we specify is the use for some no-go theorems of the sourced Bianchi identity (BI) or tadpole. This identity is an extra input which comes from 10d, but it does not appear in the 4d theory and scalar potential. It can then be interesting to see when it is needed.\\

The first observation to be made from Table \ref{tab:c} is that all computed values verify the TCC bound \eqref{TCCbound}. This was essentially observed in \cite{Bedroya:2019snp} for the no-go theorems \ref{nogo2}, \ref{nogo3} and \ref{nogo7} and we complete here the analysis by adding seven more. Interestingly, the no-go theorems \ref{nogo7} and \ref{nogo9} saturate the bound $c=\sqrt{\frac{2}{3}}$ independently of the value of $p$ (as long as $p \geq 4$) and have conditions that are field independent. The result of the no-go \ref{nogo9} does not even require the BI. This no-go was obtained in 10d in \cite{Andriot:2019wrs}, but its 4d formulation is non-trivial as it requires the derivation of a new scalar potential.

These results are at first sight remarkable: we obtained inequalities of the form \eqref{V'/V} just by combining supergravity equations. Contrary to the TCC reasoning, we have not used any quantum gravity argument, there was no need of averaging any quantity and we did not take any large distance limit. This surprising matching could be explained by interpreting supergravity equations as the classical limit of string theory (intended as quantum gravity) and by inserting this result into the wider picture of the swampland program.

\subsection{Framework, 4d scalar potentials and kinetic terms}\label{sec:framework}

We present now the framework in more detail, together with the derivation of the 4d scalar potentials and the scalar kinetic terms. Then, we will derive each of the no-go theorems in this 4d language and we will compute the corresponding values of the parameter $c$.

\subsubsection{The $(\rho,\tau,\sigma)$-potential}\label{sec:rts}

Our starting point is type IIA/B 10d supergravities with $D_p$-branes and orientifold $O_p$-planes, collectively called sources, and we follow conventions of \cite{Andriot:2016xvq}. This is the common framework where to look for classical de Sitter solutions, as reviewed in \cite{Andriot:2019wrs}. The 10d action in string frame is (up to additional Chern-Simons terms that will not contribute)
\beq
{\cal S}= \frac{1}{2 \kappa_{10}^2} \int \d^{10} x \sqrt{|g_{10}|} e^{-2\phi} \left( L_{NSNS} + L_{RR} + L_{{\rm sources}} \right) \ . \label{S10d}
\eeq
The definitions of these quantities can be found in \cite{Andriot:2016xvq}. Here, we just recall that the RR contribution in type IIA/B is explicitly given by\footnote{This 10d starting point somewhat differs from that of the seminal paper \cite{Hertzberg:2007wc}. The final 4d potential is however the same, at least in IIA. In our case, this has been obtained with a specific treatment of the 4d four-forms (or three-form gauge potentials) that is presented in appendix \ref{ap:4form} and that is crucial for a proper dimensional reduction from 10d to 4d. Our results could then be especially important in IIB.}
\beq
L_{RR}^{IIA}= -\frac{e^{2\phi}}{2} \left( |F_0|^2 + |F_2|^2 + |F_4^{10}|^2  \right) \ ,\ L_{RR}^{IIB}= -\frac{e^{2\phi}}{2} \left( |F_1|^2 + |F_3|^2 + \frac{1}{2} |F_5^{10}|^2  \right) \ ,
\eeq
where for any $q$-form $A_q$ we denote $|A_q|^2=A_{q\, M_1\dots M_q}A_{q\, N_1\dots N_q}g^{M_1 N_1} \dots g^{M_q N_q} /q!$. In the following we will simply add these two contributions together, $L_{RR} = L_{RR}^{IIA}+L_{RR}^{IIB}$, to treat both theories at once; which flux should be kept would then be obvious according to the theory considered.

We are interested here in 10d backgrounds with metric
\beq
\d s^2_{10}= \d s^2_4 + \d s^2_6 \ ,\quad \d s^2_4 = g_{\mu\nu} (x) \d x^\mu \d x^\nu \ ,\quad \d s^2_6= g_{mn} (y) \d y^m \d y^n \ ,\label{10dmetric}
\eeq
where for simplicity in this paper we do not include any warp factor. This compactification ansatz is that of a 4d maximally symmetric space-time, e.g.~de Sitter, and a 6d compact internal manifold $\mmm$. The sources, of size $p$, with $3 \leq p \leq 8$, are space-filling in 4d and wrap $p-3$ internal directions. We consider that these $p-3$ directions parallel to the sources are identified in the flat basis of the internal manifold and denoted $a_{||}$, while the $9-p$ transverse directions are denoted $a_{\bot}$. Then, the 6d metric is written as
\beq
\d s_6^2 = \delta_{ab} e^a e^b = \d s_{||}^2 + \d s_{\bot}^2 \ ,\quad  \d s_{||}^2 \equiv e^{a_{||}}{}_{m} e^{b_{||}}{}_{n} \delta_{ab} \d y^m \d y^n \ ,\quad \d s_{\bot}^2 \equiv e^{a_{\bot}}{}_{m} e^{b_{\bot}}{}_{n} \delta_{ab} \d y^m \d y^n \ ,\nn
\eeq
with the vielbeins $e^{a}{}_{m}$ depending a priori on all internal coordinates $y$. Any internal form, and in particular an internal flux $F_q$, can get decomposed on the basis $\{e^{a_{||}}\}, \{e^{a_{\bot}}\}$ and we denote $F_q^{(n)}$ the component with $n$ internal parallel flat indices. More explicitly
\beq
F_q = \frac{1}{q!} F^{(0)}_{a_{1\bot} \dots a_{q\bot}} e^{a_{1\bot}} \w \dots \w e^{a_{q\bot}} + \frac{1}{(q-1)!} F^{(1)}_{a_{1||} a_{2\bot} \dots a_{q\bot}} e^{a_{1||}} \w e^{a_{2\bot}} \w \dots \w e^{a_{q\bot}} + \dots
\eeq
and $|F_q|^2 = \sum_n |F_q^{(n)}|^2$. We refer to \cite{Andriot:2016xvq, Andriot:2018ept} for more details on these conventions.

We then consider fluctuations $\rho,\tau,\sigma > 0$ around such a background (labeled with ${}^{0}$ when necessary). These fluctuations were first introduced in \cite{Danielsson:2012et}, and most of the following derivation was made in \cite{Andriot:2018ept}. They will eventually correspond to 4d scalar fields and are defined as follows
\bea
& \d s_6^2 = \rho \left( \sigma^A (\d s_{||}^2)^0 + \sigma^B (\d s_{\bot}^2)^0 \right) \ , \ e^{a_{||}}{}_{m} = \sqrt{\rho \sigma^A}\, (e^{a_{||}}{}_{m})^0 \ ,\ e^{a_{\bot}}{}_{m} = \sqrt{\rho \sigma^B}\, (e^{a_{\bot}}{}_{m})^0 \label{fluctmetricN} \ ,\\
%& A= p-9 \ ,\ B=p-3 \ , \qquad {\rm i.e.}\ A(p-3) + B (9-p) = 0  \ ,\nn\\
& \phi = \phi^0 + \delta \phi \ ,\quad \tau = e^{- \delta \phi} \rho^{\frac{3}{2}} \ ,\ e^{\phi^0}=g_s \ ,\nn
\eea
where $A=p-9$ and $B=p-3$, i.e. $A(p-3) + B (9-p) = 0$. From such definitions, one should keep in mind that the background, i.e.~the 10d solution, is recovered by setting
\beq
\mbox{Background value:}\quad \rho=\sigma = \tau =1 \ . \label{bckgdN}
\eeq
Given these scalar fields and the compactification ansatz for the background, one can perform a dimensional reduction as described in detail in \cite{Andriot:2018ept}: one fluctuates all 10d quantities, separates 4d and 6d contributions and eventually goes to 4d Einstein frame. The final 4d action is
\beq
{\cal S}=  \int \d^4 x \sqrt{|g_{4}|} \left(\frac{M_4^2}{2} {\cal R}_{4} - \frac{1}{2} \left( (\del \hat{\tau})^2 + (\del \hat{\rho})^2 + (\del \hat{\sigma})^2  \right) - V \right) \ , \label{S4d}
\eeq
where, with respect to \cite{Andriot:2018ept}, we dropped a subscript ${}_E$ for Einstein frame and the 4d Planck mass $M_4$ differs by a factor of 2, being now given by
\beq
M_4^2 = \frac{1}{\kappa_{10}^2} \int \d^6 y \sqrt{|g_6^0|}\ g_s^{-2} \ . \label{Mp}
\eeq
We keep the Planck mass in the whole of section \ref{sec:framework}, but set it to 1 in the rest of the paper. The fields appearing in the canonically normalized kinetic terms are given by
\beq
\hat{\tau}= \sqrt{2}\, M_4\, \ln \tau \, ,\ \hat{\rho}= \sqrt{\frac{3}{2}}\, M_4\, \ln \rho \, ,\ \hat{\sigma} = \sqrt{\frac{-3AB}{2}} \, M_4\, \ln \sigma \ ,
\eeq
where $\hat{\sigma}$ was obtained in appendix B of \cite{Andriot:2019wrs}. A full computation of these kinetic terms will be provided in \cite{Andriot:2020wpp}. The scalar potential is the same as in \cite{Andriot:2018ept}, except for the $*_6 F_5$ and $F_6$ terms; those terms require a proper treatment of the 4d four-forms that we will come back to. The potential is given by
\bea
V = \frac{1}{2\kappa_{10}^2} g_s^{-2} & \int \d^6 y \sqrt{|g_6^0|} \bigg[ - \tau^{-2} \bigg( \rho^{-1} {\cal R}_6(\sigma) -\frac{1}{2} \rho^{-3} \sum_n \sigma^{-An-B(3-n)} |H^{(n)0}|^2 \bigg) \label{pot0N} \\
& \phantom{ \int \d^6 y \sqrt{|g_6^0|} \bigg[ } - g_s \tau^{-3} \rho^{\frac{p-6}{2}} \sigma^{B\frac{p-9}{2}} \frac{T_{10}^0}{p+1} \nn\\
& \phantom{ \int \d^6 y \sqrt{|g_6^0|} \bigg[ } +\frac{1}{2} g_s^2 \tau^{-4} \bigg(  \sum_{q=0}^{5} \rho^{3-q} \sum_n  \sigma^{-An-B(q-n)} |F_q^{(n)0}|^2  +  \rho^{-3} |F_6^0|^2 \bigg) \bigg] \ . \nn
\eea
The sources contribution $T_{10}$, defined in \cite{Andriot:2016xvq}, contains the charges and the $\delta$ localizing the sources in their transverse directions. A general expression for ${\cal R}_6 (\sigma)$ can be found in \cite{Andriot:2018ept}. On compact group manifolds, it can be expressed in terms of the structure constants $f^a{}_{bc}$ as
\beq
{\cal R}_6 (\sigma) = \sigma^{-B}\ {\cal R}_6^0 + (\sigma^{-A} - \sigma^{-B}) \left( {\cal R}_{||} +  {\cal R}_{||}^{\bot} \right)^0  - \frac{1}{2} (\sigma^{-2B+A} - \sigma^{-B}) |f^{{}_{||} 0}{}_{{}_{\bot} {}_{\bot}}|^2  \ ,\label{R6sigmaN3}
\eeq
where we introduce the notations
\bea
& {\cal R}_6 =  {\cal R}_{||} +  {\cal R}_{||}^{\bot}  - \frac{1}{2} |f^{{}_{||}}{}_{{}_{\bot} {}_{\bot}}|^2  - \delta^{cd}   f^{b_{\bot}}{}_{a_{||}  c_{\bot}} f^{a_{||}}{}_{ b_{\bot} d_{\bot}}  \ , \label{R6final2}\\
& 2 {\cal R}_{||}=  - \delta^{cd} f^{a_{||}}{}_{b_{||} c_{||}} f^{b_{||}}{}_{a_{||} d_{||}} - \frac{1}{2} \delta_{ad} \delta^{be} \delta^{cg} f^{a_{||}}{}_{b_{||} c_{||}} f^{d_{||}}{}_{e_{||} g_{||}} \ ,\\
& 2 {\cal R}_{||}^{\bot}=  - \delta^{cd} f^{b_{\bot}}{}_{a_{\bot} c_{||}} f^{a_{\bot}}{}_{b_{\bot} d_{||}}  - \delta^{bg}\delta^{cd}\delta_{ah} \left( f^{h_{\bot}}{}_{g_{\bot} c_{||}} f^{a_{\bot}}{}_{b_{\bot} d_{||}} +  f^{h_{\bot}}{}_{g_{||} c_{||}} f^{a_{\bot}}{}_{b_{||} d_{||}} \right) \ , \\
& |f^{{}_{||}}{}_{{}_{\bot} {}_{\bot}}|^2 = \tfrac{1}{2} \delta^{ab} \delta^{cd} \delta_{ef}  f^{e_{||}}{}_{a_{\bot} c_{\bot}} f^{f_{||}}{}_{b_{\bot} d_{\bot}} \ .
\eea
Finally, on compact group manifolds with constant fluxes, the orientifold projects out many components, leaving us only with the following $F_q^{(n)},\, H^{(n)},$
\beq
F_{6-p}^{(0)},\ F_{8-p}^{(1)},\ F_{10-p}^{(2)},\ F_{12-p}^{(3)},\ H^{(0)},\ H^{(2)} \ . \label{Fkns}
\eeq
This simplifies considerably the potential and we refer to \cite{Andriot:2018ept} for more details. Using a shorthand notation where we drop the superscript ${}^0$ and the integrals
\beq
\frac{2}{M_4^2}\, \frac{1}{2\kappa_{10}^2} g_s^{-2} \int \d^6 y \sqrt{|g_6^0|}\, |H^{(n)0}|^2 = \frac{\int \d^6 y \sqrt{|g_6^0|}\, |H^{(n)0}|^2}{ \int \d^6 y \sqrt{|g_6^0|}} \rightarrow |H^{(n)}|^2 \ , \label{simpnot}
\eeq
the useful reduced potential can be written as
\begin{empheq}[innerbox=\fbox, left=\!\!\!\!\!]{align}
\tV = \frac{2}{M_4^2}\ V = & - \tau^{-2} \bigg( \rho^{-1} {\cal R}_6(\sigma) -\frac{1}{2} \rho^{-3} \sum_n \sigma^{-An-B(3-n)} |H^{(n)}|^2 \bigg) \nn\\
&  - g_s \tau^{-3} \rho^{\frac{p-6}{2}} \sigma^{B\frac{p-9}{2}} \frac{T_{10}}{p+1}  \label{potN}\\
& +\frac{1}{2} g_s^2 \tau^{-4} \bigg(  \sum_{q=0}^{5} \rho^{3-q} \sum_n  \sigma^{-An-B(q-n)} |F_q^{(n)}|^2  +  \rho^{-3} |F_6|^2 \bigg). \nn
\end{empheq}
The generalization of this whole framework and potential to the case of intersecting sources, with scalars $\sigma_I$, can be found in \cite{Andriot:2017jhf, Andriot:2019wrs}. By freezing $\sigma = 1$, we reproduce the standard potential of the universal moduli $(\rho, \tau)$ given in \cite{Hertzberg:2007wc}, namely
\beq
\hspace{-0.2in} \tV = \frac{2}{M_4^2} V =  - \tau^{-2} \bigg( \rho^{-1} {\cal R}_6 -\frac{1}{2} \rho^{-3} |H|^2 \bigg) - g_s \tau^{-3} \rho^{\frac{p-6}{2}}  \frac{T_{10}}{p+1} +\frac{1}{2} g_s^2 \tau^{-4} \sum_{q=0}^{6} \rho^{3-q} |F_q|^2  \label{potrhotau} \ .
\eeq
This $V(\rho,\tau)$ will be enough for several no-go theorems to be derived.

In the derivation of the 4d scalar potential from the 10d theory, there is a non-trivial step consisting in taking care properly of the 4d four-forms coming from $F_4^{10}$ and $F_5^{10}$ (or equivalently of the 4d three-form gauge potentials). We describe this in detail in appendix \ref{ap:4form}. Such a step was missed in \cite{Andriot:2018ept, Andriot:2019wrs} and consequently the $*F_5$ and $F_6$ terms there in the potential are not correct. However, as discussed in appendix \ref{ap:4form}, the first derivatives $\del_{\varphi} V|_{\varphi=1}$ evaluated on the background and the 4d Ricci scalar ${\cal R}_4$ appearing in \cite{Andriot:2018ept, Andriot:2019wrs} are still correct. Indeed, they were already noticed to match with  the 10d equations of motion. Only the second derivatives would differ in these two terms, but those were barely used.\footnote{Only part of the stability island section in \cite{Andriot:2018ept} could be revisited, which amounts to minor results in any case.} In addition, $*F_5$ and $F_6$ never played any crucial role in works on classical de Sitter, as far as we know. Therefore most results of \cite{Andriot:2018ept, Andriot:2019wrs} remain correct or unaltered. Here, we will verify this by reproducing no-go theorems obtained in these papers.\\

A last, useful ingredient is the sourced Bianchi identity (BI). With notations of \cite{Andriot:2016xvq} and projected on the transverse volume to a source, this BI is written as follows
\bea
& \varepsilon_p\, \frac{T_{10}}{p+1} = (\d F_{8-p})_{\bot} - (H \w F_{6-p})_{\bot} \ , \quad \varepsilon_p=(-1)^{p+1} (-1)^{\left[\frac{9-p}{2} \right]} \label{BI1}\\
\hspace{-0.2in} \Leftrightarrow\quad &  2 \tau^{-3} \rho^{\frac{p-6}{2}} \sigma^{B \frac{p-9}{2}} g_s \frac{T_{10}}{p+1} = 2 \tau^{-3} \rho^{\frac{p-6}{2}} \sigma^{B \frac{p-9}{2}} \varepsilon_p\, g_s (\d F_{8-p})_{\bot}\nn\\
\phantom{\hspace{-0.2in} \Leftrightarrow\quad} &  \phantom{2 \tau^{-3} \rho^{\frac{p-6}{2}} \sigma^{B \frac{p-9}{2}} g_s \frac{T_{10}}{p+1} = } - 2 \varepsilon_p\, (\tau^{-1} \rho^{-\frac{3}{2}} \sigma^{-B \frac{3}{2}} H \w \tau^{-2} \rho^{\frac{p-3}{2}} \sigma^{B \frac{p-6}{2}} g_s F_{6-p})_{\bot} \nn\\
= &\  2 \tau^{-3} \rho^{\frac{p-6}{2}} \sigma^{B \frac{p-9}{2}} \varepsilon_p\, g_s (\d F_{8-p})_{\bot}  -  \left|\tau^{-1} \rho^{-\frac{3}{2}} \sigma^{-B \frac{3}{2}} *_{\bot} H^{(0)} + \varepsilon_p g_s \tau^{-2} \rho^{\frac{p-3}{2}} \sigma^{B \frac{p-6}{2}} F_{6-p}^{(0)} \right|^2 \nn\\
& + \tau^{-2} \rho^{-3} \sigma^{-3B} |H^{(0)}|^2 + g_s^2 \tau^{-4} \rho^{p-3} \sigma^{B (p-6)}  |F_{6-p}^{(0)}|^2 \ .\nn
\eea
In the rewriting of the second row, we have multiplied by scalar fields and rearranged them conveniently for the coming off-shell formulation of the no-go theorems. The background valued expression with $\rho=\tau=\sigma=1$ was already written and used e.g.~in \cite{Andriot:2016xvq}. Furthermore, we can also rewrite the term  $(\d F_{8-p})_{\bot}$ as in \cite{Andriot:2018ept}. Indeed, on compact group manifolds with an orientifold one has
\bea
& 2 \tau^{-3}  \rho^{\frac{p-6}{2}} \sigma^{B \frac{p-9}{2}} \varepsilon_p g_s(\d F_{8-p})|_{\bot}  = 2  \tau^{-3}  \rho^{\frac{p-6}{2}} \sigma^{B \frac{p-9}{2}} \varepsilon_p g_s (\d F_{8-p}^{(1)})|_{\bot} \label{BI2}\\
& = 2  \varepsilon_p g_s (\tau^{-2} \rho^{\frac{p-5}{2}} \sigma^{\frac{B-A-B(8-p)}{2}} \iota_{\del_{a_{||}}} F_{8-p}^{(1)}) \w \tau^{-1} \rho^{-\frac{1}{2}} \sigma^{\frac{A-2B}{2}} (\d e^{a_{||}})|_{\bot} \nn\\
& = - \sum_{a_{||}} \left|\tau^{-1} \rho^{-\frac{1}{2}} \sigma^{\frac{A-2B}{2}}   *_{\bot}( \d e^{a_{||}})|_{\bot} - \tau^{-2} \rho^{\frac{p-5}{2}} \sigma^{\frac{B-A-B(8-p)}{2}} \varepsilon_p g_s (\iota_{\del_{a_{||}}} F_{8-p}^{(1)} ) \right|^2 \nn\\
& + \tau^{-4} \rho^{p-5} \sigma^{B-A-B(8-p)}  g_s^2 |F_{8-p}^{(1)}|^2 + \tau^{-2} \rho^{-1} \sigma^{A-2B} |f^{{}_{||}}{}_{{}_{\bot} {}_{\bot}}|^2 \ ,\nn
\eea
where again we multiplied appropriately by scalar fields for future convenience. We will use these formulations of the sourced BI in some no-go theorems.

\subsubsection{A new scalar potential for the radius}\label{sec:radius}

We derive here the scalar potential for a different metric fluctuation: a radius $r$ to be defined below, together with a 4d dilaton $\tau$. The derivation is analogous to the one contained in the previous section, therefore we only present the main steps. We start with the same 10d action, where again the Chern-Simons and Wess-Zumino terms do not contribute because they are topological. The compactification ansatz is the same, with the metric of the compact manifold given in the flat basis as $\d s_6^2 = \delta_{ab} e^a e^b$, and we proceed as above for what concerns the sources and the fluxes. The fluctuations are defined as follows. We first choose one internal flat direction, for example $a=1$, along which we consider a fluctuation $r$. We thus split flat indices as $\{a\}=\{1,i=2 \dots 6\}$ and define
\beq
e^1{}_m= r\, (e^1{}_m)^0 \ ,\ e^i{}_m=  (e^i{}_m)^0 \ , \ \phi = \phi^0 + \delta \phi \ ,\ e^{\phi^0}=g_s \ , \ \tau = e^{- \delta \phi}\, r^{\frac{1}{2}} \ .
\eeq
We stress that the fluctuations $r$ and $\tau$ are 4d scalar fields, depending on 4d coordinates. As before, their background expectation value is
\beq
\mbox{Background value:}\quad r = \tau =1 \ . \label{bckgdr}
\eeq
The definition of $\tau$ is chosen such that, as above, the metric in 4d Einstein frame is given by $g_{E\, \mu\nu} = \tau^2 g_{\mu\nu}$. After fluctuating 10d quantities in the 10d initial action, separating between 4d and 6d and going to 4d Einstein frame, one reaches eventually the following action
\beq
{\cal S}=  \int \d^4 x \sqrt{|g_{4}|} \left(\frac{M_4^2}{2} {\cal R}_{4} - \frac{1}{2} \left( (\del \hat{\tau})^2 + (\del \hat{r})^2 \right) - V (r,\tau) \right) \ , \label{S4dr}
\eeq
where we drop a subscript ${}_E$ for Einstein frame and where the Planck mass is defined as in \eqref{Mp}. The canonical fields $\hat{\tau}$ and $\hat{r}$ are given by
\beq
\hat{\tau}= \sqrt{2}\, M_4\, \ln \tau \, ,\ \hat{r}= \, M_4\, \ln r \ . \label{cantaur}
\eeq
Their kinetic terms are computed as usual from derivative terms, namely from ${\cal R}_{10} + 4 (\del \phi)^2$; we will come back to it.\footnote{\label{foot:linkrtsr}In the framework of section \ref{sec:rts}, one may specify to the following subcase: $p=4$, $\rho=\sigma^{-B}=\sigma^{-1}$. In that case, the fluctuations considered there match the ones considered here, with $r=\sigma^{-3}=\rho^3$. One verifies then that $\tau$ is the same, as well as the kinetic terms. These are interesting cross-checks. Note that here formulas remain valid for any $p\geq 4$.} The scalar potential is obtained by fluctuating all internal quantities, as well as by going on-shell with respect to the four-form fluxes. As we did previously for $\sigma$, we need to distinguish the internal flux components which are along direction $1$ and those which are not:
\beq
F_q = \frac{1}{q!} F^{(0)}_{i_{1} \dots i_{q}} e^{i_{1}} \w \dots \w e^{i_{q}} + \frac{1}{(q-1)!} F^{(1)}_{1\, i_{2} \dots i_{q}} e^{1} \w e^{i_{2}} \w \dots \w e^{i_{q}} \ .
\eeq
In particular, we get $|F_q|^2 = |F_q^{(0)}|^2 + |F_q^{(1)}|^2 = |F_q^{(0)0}|^2 + r^{-2} |F_q^{(1)0}|^2$. Similarly for the source contribution, coming from the DBI action, one has to distinguish whether the direction 1 is parallel or transverse to the source. This is done with the symbols $\delta_1^{||}$ and $\delta_1^{\bot}$. The 4d potential in \eqref{S4dr} is therefore given by
\bea
V(r,\tau) = \frac{1}{2\kappa_{10}^2} g_s^{-2}  \int \d^6 y \sqrt{|g_6^0|}  \bigg[ & - \tau^{-2} \left( {\cal R}_6(r) -\frac{1}{2} ( |H^{(0)0}|^2 + r^{-2} |H^{(1)0}|^2)\right) \label{potrtau0}\\
 & - g_s \tau^{-3}  (\delta_1^{||} r^{\frac{1}{2}} + \delta_1^{\bot} r^{-\frac{1}{2}})  \frac{T_{10}^0}{p+1} \nn\\
 & + \frac{1}{2} g_s^2 \tau^{-4} \left( \sum_{q=0}^{5} ( r |F_q^{(0)0}|^2 + r^{-1} |F_q^{(1)0}|^2) +   r^{-1} |F_6^{0}|^2 \right) \bigg] \ . \nn
\eea
The $F_5$ and $F_6$ terms are obtained after taking care of four-form fluxes, a point we detail in appendix \ref{ap:4formr}. Finally, ${\cal R}_6(r)$ is obtained by fluctuating
\bea
{\cal R}_6 = \delta^{ab} {\cal R}_{ab} &  = -\frac{1}{2} \delta^{cd} \left(f^b{}_{ac} f^a{}_{bd} + \delta^{ef} \, \, \delta_{ab} \, f^a{}_{ec} \, f^b{}_{fd} - \frac{1}{2} \delta^{eh} \,\delta^{fg} \, \delta_{ac} \delta_{bd} \, f^a{}_{ef} \, f^b{}_{hg} \right)\\
& =-\frac{1}{2}  \delta^{cd} f^b{}_{ac} \, f^a{}_{bd} - \frac{1}{4} \delta^{eb} \,\delta^{fc} \, \delta_{ga} \, f^a{}_{ef} \, f^g{}_{bc} \ ,\nn
\eea
where we specialized to compact group manifolds, for which $f^b{}_{ac}$ are constant and $f^a{}_{ac}=0$. After some rewriting, we obtain
\beq
{\cal R}_6(r) = {\cal R}_6^0 + (r^{-2}-1) {\cal R}_{11}^0 + \frac{1}{4} (2- r^2 -r^{-2}) \delta^{ik} \,\delta^{jl} \, {f^{1}{}_{ij}}^0 \, {f^{1}{}_{kl}}^0  \ .
\eeq
One can easily verify that ${\cal R}_6(r=1) = {\cal R}_6^0$ as it should in the background. Introducing the same simplified notations as in \eqref{simpnot}, we obtain eventually the reduced potential
\begin{empheq}[innerbox=\fbox, left=\!\!\!\!\!]{align}
\tV = \frac{2}{M_4^2}\ V = & - \tau^{-2} \left( {\cal R}_6(r) -\frac{1}{2} ( |H^{(0)}|^2 + r^{-2} |H^{(1)}|^2)\right) \nn\\
&  - g_s \tau^{-3}  (\delta_1^{||} r^{\frac{1}{2}} + \delta_1^{\bot} r^{-\frac{1}{2}})  \frac{T_{10}}{p+1} \label{potredrt}\\
& + \frac{1}{2} g_s^2 \tau^{-4} \left( \sum_{q=0}^{5} ( r |F_q^{(0)}|^2 + r^{-1} |F_q^{(1)}|^2) +   r^{-1} |F_6|^2 \right). \nn
\end{empheq}

To compute the kinetic terms, we can specify without loss of generality to the following 10d metric
\beq
\d s_{10}^2 = \tau^{-2}(x) \eta_{\mu \nu} \d x^{\mu} \d x^{\nu} + r^2 (x) (\d y^1)^2 + \delta_{\ij} \d y^i \d y^j \ ,
\eeq
where $\eta_{\mu \nu}$ is the 4d Einstein frame flat metric, used to raise and lower indices in the expressions below, as well as in the squares. One then computes
\bea
& {\cal R}_{10} = \tau^2 \del_{\mu}\left( 6\, \del^{\mu} \ln \tau - 2\, \del^{\mu} \ln r \right) - 6 (\del \tau)^2 - 2 \tau^2 (\del \ln r)^2 + 4 \tau^2\, \del_{\mu} \ln r \, \del^{\mu} \ln \tau \ ,\label{R10}\\
& 4 |\del \phi|^2 = 4(\del \tau)^2 + \tau^2 (\del \ln r)^2 - 4 \tau^2\, \del_{\mu} \ln r\, \del^{\mu} \ln \tau \ ,
\eea
where $|\del \phi|^2$ is the 10d string frame square. The kinetic part of the 10d action \eqref{S10d} contains the terms
\bea
{\cal S}_{kin} &= \frac{1}{2 \kappa_{10}^2} \int \d^{10} x \sqrt{|g_{10}|} e^{-2\phi} \left( {\cal R}_{10} + 4 |\del \phi|^2 \right) \\
&=\int \d^{4} x \sqrt{|g_{4}|} \, \frac{g_s^{-2}}{2 \kappa_{10}^2} \int \d^{6} y \sqrt{|g_{6}^0|}\, \tau^{-2} \left( {\cal R}_{10} + 4 |\del \phi|^2 \right) \ ,\nn
\eea
where in the second line we moved to 4d Einstein frame. The factor $\tau^{-2}$ makes the first term in ${\cal R}_{10}$ in \eqref{R10} a total derivative that we drop when substituting it into the action, leaving us with
\beq
{\cal S}_{kin} = \int \d^{4} x \sqrt{|g_{4}|} \, \frac{M_4^2}{2}  \left( - 2 \tau^{-2} (\del \tau)^2 - (\del \ln r)^2   \right) \ .\label{Skinfinal}
\eeq
Using a more general metric, we would have obtained terms with ${\cal R}_4$, ${\cal R}_6$ and $\nabla_{\mu}$. However, we can easily see that the expression \eqref{Skinfinal} is sufficient to reproduce the kinetic terms \eqref{S4dr} for the canonical fields \eqref{cantaur}.

\subsection{No-go theorems in 4d}\label{sec:nogolist}

Having constructed the scalar potentials \eqref{potN} and \eqref{potredrt}  of the four-dimensional theory, we now take a step forward and reproduce in 4d the known no-go theorems on de Sitter solutions, by considering appropriate combinations of the scalar potential and its first derivatives. These combinations can be shown to be negative, provided that some condition holds, which is the assumption of the no-go theorem. In other words, we will obtain inequalities of the type
\beq
a\, V + \sum_{i=1}^3 b_i\, \del_{\hat{\varphi}^i} V \leq 0 \ ,\ a>0\ ,\ \exists\, b_i\neq 0 \ ,\label{combinogo}
\eeq
where $\hat{\varphi}^i$ are the canonically normalized scalar fields, i.e.~for which $g_{ij}=\delta_{ij}$; we will not consider more than three fields here. Such an inequality clearly forbids a de Sitter extremum, so it provides a no-go. Moreover, it will also play a crucial role in section \ref{sec:checkmap}. From \eqref{combinogo}, one can then obtain an inequality of the form \eqref{V'/V} and deduce a value for $c$. To that end, we follow section 3.1 of \cite{Andriot:2019wrs} and the method of \cite{Hertzberg:2007wc}: we get a lower bound on $|\nabla V|/V$ using the inequality \eqref{combinogo} and minimizing a function. This bound $c$ is given by
\beq
c^2 = \frac{a^2}{b_1^2 + b_2^2 +b_3^2} \ .\label{cformula}
\eeq
This is how we proceed in the following for each no-go theorem. In this section we set $M_4 =1$ for simplicity.

We first consider no-go theorems \ref{nogo1} - \ref{nogo4} that only require the two-field potential $V(\rho,\tau)$ \eqref{potrhotau}. They are thus valid for both parallel or intersecting sources.\\
\begin{enumerate}[label=\textbf{\arabic*.}]
  \item \label{nogo1}{\bf Maldacena-Nu\~nez}
\end{enumerate}
We reproduce this no-go theorem \cite{Maldacena:2000mw} through the following combination
\bea
& 2 \tV + \tau \del_{\tau} \tV = \tau^{-3} \rho^{\frac{p-6}{2}} g_s \frac{T_{10}}{p+1} - g_s^2  \tau^{-4} \sum_{q=0}^{6} \rho^{3-q} |F_q|^2  \\
\Rightarrow\quad & 2 V + \sqrt{2}\, \del_{\hta} V \leq 0 \qquad {\rm if}\ T_{10} \leq 0  \\
\Rightarrow\quad &  c^2=2 \ .
\eea

\begin{enumerate}[label=\textbf{\arabic*.},resume]
  \item \label{nogo2}{\bf No-go for $p=7,8$, or $p=4,5,6$ with $F_{6-p}=0$}
\end{enumerate}
This no-go was derived in 4d in \cite{Wrase:2010ew} (see also \cite{Shiu:2011zt} and references therein) and in 10d in \cite{Andriot:2016xvq}. As pointed-out in \cite{Andriot:2018ept}, for $p=4$ sources, the $O_4$ projection combined with the $F_0$ BI imply $F_0=0$. So one only needs to consider $F_2$ for $p=4$. One shows explicitly
\bea
& 2 (p-3) \tV + (p-4) \tau \del_{\tau} \tV + 2 \rho \del_{\rho} \tV \\
 =\ & - \tau^{-2} \rho^{-3} 2 |H|^2 + g_s^2 \tau^{-4} \sum_{q=0}^{6} \rho^{3-q} (8-p-q) |F_q|^2  \nn \\
\Rightarrow\quad &  2 (p-3) V + (p-4) \sqrt{2}\, \del_{\hta} V + \sqrt{6}\, \del_{\hr} V \leq 0 \quad \mbox{if $p=7,8$, or $p=4,5,6$ $\&$ $F_{6-p}=0$} \\
\Rightarrow\quad & c^2=\frac{2(p-3)^2}{3+(p-4)^2}   \ .
\eea
We get $c^2=\frac{2}{3}$ for $p=4$ and higher otherwise. Although the equations are valid for $p=3$, they do not provide an appropriate inequality since the $V$ term vanishes. This $c$ value was obtained already in \cite{Wrase:2010ew}. It matches the values given for $\R_6 <0$ in table 1 of \cite{Obied:2018sgi}.

\begin{enumerate}[label=\textbf{\arabic*.},resume]
  \item \label{nogo3}{\bf Positive or vanishing curvature}
\end{enumerate}
This no-go was derived in 4d in \cite{Wrase:2010ew} (see also references therein) and in 10d in \cite{Andriot:2016xvq}. As above, we use the argument of \cite{Andriot:2018ept} to conclude that $F_0=0$ for $p=4$. One shows explicitly
\bea
& 2 (p+3) \tV + p\, \tau \del_{\tau} \tV + 2 \rho \del_{\rho} \tV \\
 =\ & - \tau^{-2} \rho^{-1} 4 \R_6 + g_s^2 \tau^{-4} \sum_{q=0}^{6} \rho^{3-q} (6-p-q) |F_q|^2 \nn \\
\Rightarrow\quad &  2 (p+3) V + p \sqrt{2}\, \del_{\hta} V + \sqrt{6}\, \del_{\hr} V \leq 0 \quad \mbox{if $\R_6 \geq 0$ and $p \geq 4$} \\
\Rightarrow\quad & c^2=\frac{2(p+3)^2}{3+p^2} > 1  \ .
\eea
This can be extended to $p=3$ with $F_1=0$. This $c$ value was given in \cite{Wrase:2010ew} already and matches those for $\R_6 \geq 0$ in \cite{Obied:2018sgi}. The case $p=6$ is the one discussed in the seminal paper \cite{Hertzberg:2007wc}.

\pagebreak

\begin{enumerate}[label=\textbf{\arabic*.},resume]
  \item \label{nogo4}{\bf No-go for $p=3$}
\end{enumerate}
This no-go was derived in 10d in \cite{Blaback:2010sj, Andriot:2016xvq}. This is the first instance where we use additionally the sourced Bianchi identity \eqref{BI1}, where we can for now set $\sigma=1$. One shows
\bea
& 2 (p+1) \tV + p\, \tau \del_{\tau} \tV + 2 \rho \del_{\rho} \tV \\
 =\ & \tau^{-3} \rho^{\frac{p-6}{2}} 4 g_s \frac{T_{10}}{p+1}  - \tau^{-2} \rho^{-3} 2 |H|^2 \nn\\
 & + g_s^2  \tau^{-4} \sum_{q=0}^{6} \rho^{3-q} (4-p-q) |F_q|^2  \nn\\
 =\ &  4 \tau^{-3} \rho^{\frac{p-6}{2}} \varepsilon_p\, g_s (\d F_{8-p})_{\bot}  - 2 \left|\tau^{-1} \rho^{-\frac{3}{2}} *_{\bot} H^{(0)} + \varepsilon_p g_s \tau^{-2} \rho^{\frac{p-3}{2}} F_{6-p}^{(0)} \right|^2\nn\\
&  - 2 \tau^{-2} \rho^{-3} (|H|^2- |H^{(0)}|^2) - 2 g_s^2 \tau^{-4} \rho^{p-3} (|F_{6-p}|^2 -|F_{6-p}^{(0)}|^2 )\nn\\
 & + g_s^2  \tau^{-4} \sum_{q=0,\, q\neq 6-p}^{6} \rho^{3-q} (4-p-q) |F_q|^2 \ .\nn\\
\eea
All terms on the right-hand side are negative, except the first one in $(\d F_{8-p})_{\bot}$ that has a priori no definite sign. By the notation $(\d F_{8-p})_{\bot}$, we mean the coefficient of $\d F_{8-p}$ along the transverse directions ${\rm vol}_{\bot}$. We recall that each term appearing above is to be read with the simplified notation \eqref{simpnot}, where an integral and a vacuum value are understood, in particular
\beq
\frac{\int \d^6 y \sqrt{|g_6^0|} (\d F_{8-p}^{0})_{\bot} }{\int \d^6 y \sqrt{|g_6^0|}} \leftrightarrow (\d F_{8-p})_{\bot} \ . \label{notation}
\eeq
For $p=3$, the whole 6d space is transverse, so $\int \d^6 y \sqrt{|g_6^0|} (\d F_{5})_{\bot} = \int \d F_{5} = 0 $ since this is a total derivative over a compact space. So the first term vanishes for $p=3$ leaving us with a no-go. More generally, we deduce from above
\bea
\Rightarrow\quad & 2 (p+1) V + p \sqrt{2}\, \del_{\hta} V + \sqrt{6}\, \del_{\hr} V \leq 0 \quad \mbox{if $\varepsilon_p\, (\d F_{8-p})_{\bot} \leq 0 $} \\
\Rightarrow\quad & c^2=\frac{2(p+1)^2}{3+p^2}   \ .
\eea
For $p=3$ we obtain $c^2 = \frac{8}{3}$, and smaller values (greater than 2) for higher $p$.\\

For the no-go theorems \ref{nogo5}-\ref{nogo8}, we will use the complete potential $V(\rho,\tau,\sigma)$ \eqref{potN}, i.e.~we consider also the dependence on $\sigma$. This means we now focus on parallel sources only. For simplicity, we restrict ourselves to compact group manifolds with constant fluxes, even though several no-go theorems can be extended beyond this case \cite{Andriot:2016xvq, Andriot:2017jhf}. These assumptions provide many simplifications in the potential, as detailed in \eqref{R6sigmaN3} - \eqref{Fkns}.\\

\pagebreak

\begin{enumerate}[label=\textbf{\arabic*.},resume]
  \item \label{nogo5}{\bf  The internal parallel Einstein trace}
\end{enumerate}
With the above information and restrictions, we use the potential \eqref{potN} to show
\bea
& 3B \tV + \frac{3}{2} B \tau \del_{\tau} \tV + B \rho \del_{\rho} \tV - \sigma \del_{\sigma} \tV \\
 =\ & (A-B) \Bigg( - \tau^{-2}  \rho^{-1} \sigma^{-A} ( {\cal R}_{||} + {\cal R}_{||}^{\bot}) - \frac{1}{2} \tau^{-2}  \rho^{-1} \sigma^{-2B+A} |f^{{}_{||}}{}_{{}_{\bot} {}_{\bot}}|^2 \nn\\
&\phantom{(A-B) \Bigg(} + \frac{1}{2}  \sum_n n \Big(  \tau^{-2} \rho^{-3} \sigma^{-An-B(3-n)} |H^{(n)}|^2  +  g_s^2 \tau^{-4} \sum_{q=0}^5 \rho^{3-q} \sigma^{-An-B(q-n)} |F_q^{(n)}|^2 \Big) \Bigg) \nn\\
& - 3 B g_s^2  \tau^{-4} \rho^{-3} |F_6|^2  \ ,\nn
\eea
and we recall $A-B<0$. We deduce, with the usual values $B=p-3$, $A=p-9$,
\bea
\quad & 3B V + \frac{3}{2} B \sqrt{2}\, \del_{\hta} V + B \sqrt{\frac{3}{2}}\, \del_{\hr} V - \sqrt{\frac{-3AB}{2}} \del_{\hat{\sigma}} \tV  \leq 0 \ ,\\
& \phantom{3B V + 32 B \sqrt{2}\, \del_{\hta} V + B 32\, \del_{\hr} V - 3AB2 \del_{\hat{\sigma}} \tV}\quad  \mbox{if $  {\cal R}_{||} + {\cal R}_{||}^{\bot} + \frac{1}{2} \sigma^{2(A-B)} |f^{{}_{||}}{}_{{}_{\bot} {}_{\bot}}|^2 \leq 0 $} \nn \\
\Rightarrow\quad & c^2=\frac{6B}{4B-A}=\frac{2(p-3)}{p-1}   \ .
\eea
We get $c^2 = \frac{2}{3}$ for $p=4$ and higher otherwise; again one cannot apply this inequality to $p=3$ since the term proportional to $V$ would have a vanishing coefficient.

We face for the first time the phenomenon that the condition for a no-go becomes field dependent off-shell. This no-go was first derived in 10d in \cite{Andriot:2016xvq}, where the condition was necessarily on-shell, so field independent. The off-shell field dependence puts such a no-go on weaker grounds, since one rather asks for generic assumptions in a no-go theorem. However we notice that the corresponding requirement for a de Sitter solution can become field independent, namely $|f^{{}_{||}}{}_{{}_{\bot} {}_{\bot}}|^2 >0$, in the case where ${\cal R}_{||} + {\cal R}_{||}^{\bot} \leq 0$, which often happens. In this perspective, such a condition remains interesting.

\begin{enumerate}[label=\textbf{\arabic*.},resume]
  \item \label{nogo6}{\bf  The linear combination}
\end{enumerate}
We first consider the following combination
\bea
& 8 \tV + 3 \tau \del_{\tau} \tV + \frac{2}{3} (6-p) \rho \del_{\rho} \tV + \frac{2}{3} \sigma \del_{\sigma} \tV \\
 =&\ 4  \tau^{-3}  \rho^{\frac{p-6}{2}} \sigma^{B \frac{p-9}{2}} g_s\frac{T_{10}}{p+1} -2  \tau^{-2}  \rho^{-1} \sigma^{-2B+A} |f^{{}_{||}}{}_{{}_{\bot} {}_{\bot}}|^2 \nn\\
&  -4  \tau^{-2}  \rho^{-1} \sigma^{-A} ( {\cal R}_{||} + {\cal R}_{||}^{\bot}) + 2  \sum_n  \tau^{-2} \rho^{-3} \sigma^{-An-B(3-n)} (n - 1 ) |H^{(n)}|^2 \nn\\
& - g_s^2 \tau^{-4} \sum_{q=0}^5 \rho^{3-q} \sigma^{-An-B(q-n)} (p+q-4 -2n) |F_q^{(n)}|^2 -  g_s^2 \tau^{-4} \rho^{-3} (8-p) |F_6|^2 \nn\ ,
\eea
using $A-B=-6$. We now use the sourced BI \eqref{BI1} combined with \eqref{BI2} to replace $T_{10}$ in the above combination. We get
\bea
& 8 \tV + 3 \tau \del_{\tau} \tV + \frac{2}{3} (6-p) \rho \del_{\rho} \tV + \frac{2}{3} \sigma \del_{\sigma} \tV \\
 = &\ - 2 \left|\tau^{-1} \rho^{-\frac{3}{2}} \sigma^{-B \frac{3}{2}} *_{\bot} H^{(0)} + \varepsilon_p g_s \tau^{-2} \rho^{\frac{p-3}{2}} \sigma^{B \frac{p-6}{2}} F_{6-p}^{(0)} \right|^2 \nn\\
&  - 2 \sum_{a_{||}} \left|\tau^{-1} \rho^{-\frac{1}{2}} \sigma^{\frac{A-2B}{2}}   *_{\bot}( \d e^{a_{||}})|_{\bot} - \tau^{-2} \rho^{\frac{p-5}{2}} \sigma^{\frac{B-A-B(8-p)}{2}} \varepsilon_p g_s (\iota_{\del_{a_{||}}} F_{8-p}^{(1)} ) \right|^2 \nn\\
& + 2  \tau^{-2} \rho^{-3} \sigma^{-2A-B} |H^{(2)}|^2  -4   \tau^{-2}  \rho^{-1} \sigma^{-A} ( {\cal R}_{||} + {\cal R}_{||}^{\bot}) \nn\\
& + 2 \tau^{-4} \rho^{p-5} \sigma^{B-A-B(8-p)}  g_s^2 |F_{8-p}^{(1)}|^2 +  2 g_s^2 \tau^{-4} \rho^{p-3} \sigma^{B (p-6)}  |F_{6-p}^{(0)}|^2 \nn\\
& - g_s^2 \tau^{-4} \sum_{q=0}^5 \rho^{3-q} \sigma^{-An-B(q-n)} (p+q-4-2n) |F_q^{(n)}|^2 -  g_s^2 \tau^{-4} \rho^{-3} (8-p) |F_6|^2  \nn\ .
\eea
One can show that the last two lines, i.e.~the RR fluxes contributions, are negative. We deduce
\bea
\quad & 8 V + 3\sqrt{2}\, \del_{\hta} V + \sqrt{\frac{2}{3}} (6-p) \, \del_{\hr} V + \sqrt{\frac{-2AB}{3}} \del_{\hat{\sigma}} \tV  \leq 0 \ , \\
& \phantom{8 V + 3\sqrt{2}\, \del_{\hta} V + 23 (6-p) \, \del_{\hr} V + 2AB3 \del_{\hat{\sigma}} \tV } \quad \mbox{if $ - 2   \rho^{2} \sigma^{A+B} ( {\cal R}_{||} + {\cal R}_{||}^{\bot}) +  |H^{(2)}|^2   \leq 0 $ } \nn \\
\Rightarrow\quad & c^2=\frac{8}{3} \ .
\eea
This no-go theorem was first derived on-shell in 10d in \cite{Andriot:2016xvq}. It was shown that the linear combination $ - 2  ( {\cal R}_{||} + {\cal R}_{||}^{\bot}) +  |H^{(2)}|^2$ is a special one, being in some cases T-duality invariant. Now we face again the phenomenon that it becomes a field dependent quantity off-shell, which weakens the generality of this no-go. However, one of the two terms could be zero (no constraint forbids it so far), in which case one recovers a field independent condition. Let us also recall that this no-go theorem can be combined with the previous one (where one can include an additional $ |H^{(2)}|^2$ term), leading to the requirement $|f^{{}_{||}}{}_{{}_{\bot} {}_{\bot}}|^2 \neq 0$ for de Sitter. Finally, we add that this no-go is the $p\geq 4$ generalization of the no-go \ref{nogo4}, as the $c$ value and the use of the BI suggest; in particular for $p=3$ the linear combination vanishes.

\begin{enumerate}[label=\textbf{\arabic*.},resume]
  \item \label{nogo7}{\bf The requirement $\lambda > 0$}
\end{enumerate}
Given the requirement for de Sitter to have $|f^{{}_{||}}{}_{{}_{\bot} {}_{\bot}}|^2 \neq 0$, we can introduce the parameter
\beq
\lambda= - \frac{\delta^{cd}   f^{b_{\bot}}{}_{a_{||}  c_{\bot}} f^{a_{||}}{}_{ b_{\bot} d_{\bot}}}{ |f^{{}_{||}}{}_{{}_{\bot} {}_{\bot}}|^2 } \ . \label{deflambda}
\eeq
We obtained in \cite{Andriot:2018ept} the following no-go theorem on-shell, that we generalize here off-shell
\bea
&  2 \tV + \frac{3}{2} \tau \del_{\tau} \tV + \frac{A+B}{A-B} \rho \del_{\rho} \tV + \frac{2}{B-A} \sigma \del_{\sigma} \tV\\
= & - 2\, \delta^{cd}   f^{b_{\bot}}{}_{a_{||}  c_{\bot}} f^{a_{||}}{}_{ b_{\bot} d_{\bot}} \tau^{-2} \rho^{-1} \sigma^{3-p} \nn\\
& - 2  \left|\tau^{-1} \rho^{-\frac{3}{2}} \sigma^{-B \frac{3}{2}} *_{\bot} H^{(0)} + \varepsilon_p g_s \tau^{-2} \rho^{\frac{p-3}{2}} \sigma^{B \frac{p-6}{2}} F_{6-p}^{(0)} \right|^2  \nn\\
& - 2 \sum_{a_{||}} \left|\tau^{-1} \rho^{-\frac{1}{2}} \sigma^{\frac{A-2B}{2}}   *_{\bot}( \d e^{a_{||}})|_{\bot} - \tau^{-2} \rho^{\frac{p-5}{2}} \sigma^{\frac{B-A-B(8-p)}{2}} \varepsilon_p g_s (\iota_{\del_{a_{||}}} F_{8-p}^{(1)} ) \right|^2 \nn\\
& + 2 \tau^{-4} g_s^2 ( \rho^{p-5} \sigma^{B-A-B(8-p)}   |F_{8-p}^{(1)}|^2 +  \rho^{p-3} \sigma^{B (p-6)}  |F_{6-p}^{(0)}|^2) \nn\\
& - \frac{1}{2} g_s^2 \tau^{-4} \sum_{q=0}^5 \rho^{3-q} \sigma^{-An-B(q-n)} (p+q-2-2n) |F_q^{(n)}|^2 - \frac{1}{2} g_s^2 \tau^{-4} \rho^{-3} (10-p) |F_6|^2 \nn\ ,
\eea
where we used the sourced BI \eqref{BI1}, \eqref{BI2}. Again, the RR fluxes terms are negative. We deduce
\bea
\Rightarrow\quad & 2 V + \frac{3}{2}\sqrt{2}\, \del_{\hta} V + \frac{A+B}{A-B} \sqrt{\frac{3}{2}} \, \del_{\hr} V + \frac{1}{B-A} \sqrt{-6AB} \,  \del_{\hat{\sigma}} \tV  \leq 0 \quad \mbox{if $\lambda \leq 0 $} \nn \\
\Rightarrow\quad & c^2=\frac{2}{3} \ , \qquad \mbox{as computed on-shell in section 3.1 of \cite{Andriot:2019wrs}} \, .
\eea
Remarkably, we get the same value of $c$ value for all of the allowed $p$. We implicitly restricted to $p\geq 4$ given the presence of parallel directions; it would be interesting to see if this inequality could survive for $p=3$.

\begin{enumerate}[label=\textbf{\arabic*.},resume]
  \item \label{nogo8}{\bf The (on-shell) requirement $\lambda < 1$}
\end{enumerate}
We obtained in \cite{Andriot:2018ept} the following no-go theorem on-shell, that we generalize here off-shell
\bea
&  2 \tV + \frac{1}{2}\frac{A-5B}{A-3B} \tau \del_{\tau} \tV + \frac{1}{3} \rho \del_{\rho} \tV + \frac{2}{3(A-3B)} \sigma \del_{\sigma} \tV\\
= & -\frac{2}{p}\, \tau^{-2} \rho^{-1} \sigma^{-3-p} |f^{{}_{||}}{}_{{}_{\bot} {}_{\bot}}|^2 (\lambda \sigma^{6}-1 )  - \frac{2}{p}\, \tau^{-2} \rho^{-3} \sigma^{-(B+2A)} |H^{(2)}|^2 \nn\\
& - \frac{1}{2p} g_s^2 \tau^{-4} \sum_{q=0}^5 \rho^{3-q} \sigma^{-An-B(q-n)} (p+q-6+2n) |F_q^{(n)}|^2 - \frac{3}{2p} g_s^2 \tau^{-4} \rho^{-3} (p-2) |F_6|^2 \nn\ .
\eea
All fluxes contributions are negative in our setting, so we deduce
\bea
\Rightarrow\quad & 2 V + \frac{\sqrt{2}}{2}\frac{A-5B}{A-3B} \, \del_{\hta} V + \sqrt{\frac{1}{6}} \, \del_{\hr} V + \frac{1}{A-3B} \sqrt{\frac{-2AB}{3}} \,  \del_{\hat{\sigma}} V  \leq 0 \quad \mbox{if $\lambda \geq \sigma^{-6} $} \nn \\
\Rightarrow\quad & c^2=6\frac{A-3B}{A-7B} = \frac{2p}{p-2} > 1 \ , \ \ \ \mbox{as computed on-shell in section 3.1 of \cite{Andriot:2019wrs}} \, .
\eea
Unfortunately, the condition for the no-go has now become field dependent, and contrary to previous no-gos, we do not see any interesting subcase where the field dependence can disappear. This no-go theorem may therefore not make sense off-shell.\\

All no-go theorems considered so far got generalized to some extent in \cite{Andriot:2017jhf, Andriot:2019wrs} to cases with intersecting sources. We however do not find it relevant to recall these generalizations: either they eventually lead to the same inequalities on the potential (up to generalizations of the assumptions), or there exists no good generalization. Further no-go theorems were obtained in \cite{Andriot:2019wrs} in both the case of parallel or intersecting sources by using the $H$-flux e.o.m. The latter is however complicated to obtain generically with our scalar potential, so we refrain from translating those here.\\

\begin{enumerate}[label=\textbf{\arabic*.},resume]
  \item \label{nogo9}{\bf  The internal parallel Einstein equation}
\end{enumerate}
We aim here at reproducing in 4d and off-shell the 10d no-go theorem on parallel internal Einstein equations discussed in section 3.2 of \cite{Andriot:2019wrs}. It requires to use the new scalar potential \eqref{potredrt} derived in section \ref{sec:radius} that depends on a new scalar field, the radius $r$, as well as the 4d dilaton $\tau$. We now consider the case in which the radius direction $1$ is parallel to the sources, meaning $\delta_1^{\bot}=0$. This requires in particular $p \geq 4$. From the potential \eqref{potredrt}, one obtains
\bea
2 \tV + \tau \del_{\tau} \tV + 2 r\del_r \tV = &  - 2 \tau^{-2} \left( r\del_r {\cal R}_6(r) + r^{-2} |H^{(1)}|^2 \right) \label{nogor1} \\
& -2 \tau^{-4}  r^{-1} g_s^2 \left(\sum_{q=0}^{5} |F_q^{(1)}|^2 +   |F_6|^2 \right) \nn \ ,
\eea
where all terms in the right-hand side are negative except the first one. One has
\bea
-2 r\del_r {\cal R}_6(r)  &=  4 r^{-2} {\cal R}_{11} +  (r^2- r^{-2}) \delta^{ik} \,\delta^{jl} \, {f^{1}{}_{ij}} \, {f^{1}{}_{kl}}  \\
& = -2 r^{-2}  \left(f^j{}_{i1} f^i{}_{j1} + \delta^{ij} \, \, \delta_{kl} \, f^k{}_{i1} \, f^l{}_{j1} \right) +  r^2 \delta^{ik} \,\delta^{jl} \, {f^{1}{}_{ij}} \, {f^{1}{}_{kl}} \ ,\nn
\eea
and we recall that
\beq
f^j{}_{i1} f^i{}_{j1} + \delta^{ij} \, \, \delta_{kl} \, f^k{}_{i1} \, f^l{}_{j1} = \frac{1}{2} (\delta_{jk} \, f^k{}_{i1} + \delta_{il} \, f^l{}_{j1} )^2 \ ,
\eeq
where the square is given by contractions with $\delta^{..}$, without a factor. So the square $\delta^{ik} \,\delta^{jl} \, {f^{1}{}_{ij}} \, {f^{1}{}_{kl}}$ is the only possible positive contribution. We deduce that a de Sitter solution requires at least some $i,j$ such that $f^{1}{}_{ij} \neq 0$. This reproduces the no-go theorem of \cite{Andriot:2019wrs}, since $1$ stands in fact for any parallel direction. Note that on-shell, one can also require ${\cal R}_{11} > 0$ as in \cite{Andriot:2019wrs}, but this gets modified off-shell. Using the canonically normalized fields \eqref{cantaur}, we deduce
\bea
\Rightarrow\quad &  2 V + \sqrt{2}\, \del_{\hta} V + 2\, \del_{\hat{r}} V \leq 0 \quad \mbox{if $f^{1}{}_{ij} = 0$ for any parallel direction $1$ and all $i,j$} \\
\Rightarrow\quad & c^2=\frac{2}{3} \ .
\eea
Remarkably, this value is independent of $p \geq 4$.

\begin{enumerate}[label=\textbf{\arabic*.},resume]
  \item \label{nogo10}{\bf Heterotic at order $({\alpha'})^0$}
\end{enumerate}

De Sitter string backgrounds at all orders in $\alpha'$ and tree-level in $g_s$ have been completely excluded in heterotic string \cite{Green:2011cn, Gautason:2012tb, Kutasov:2015eba, Quigley:2015jia}. At order $\alpha'$, a simple reason is the universal scaling of the action with the dilaton \cite{Gautason:2012tb}: this leads to the same linear combination as the one of no-go \ref{nogo1}, which excludes de Sitter and gives $c=\sqrt{2}$. Interestingly, there is another combination one can consider at order $({\alpha'})^0$, forgetting about the dilaton but considering the volume. At order $({\alpha'})^0$, the (bosonic) effective action for heterotic string is just the NSNS piece of \eqref{S10d}: the $(\rho,\tau)$ potential \eqref{potrhotau} is then reduced to only the ${\cal R}_6$ and $H$-flux terms. From this we get the following identities
\bea
& \tV + \rho \del_{\rho} \tV = - \tau^{-2} \rho^{-3}  |H|^2  \\
\Rightarrow\quad & V + \sqrt{\frac{3}{2}}\, \del_{\hr} V \leq 0  \label{nogo10ineq} \\
\Rightarrow\quad &  c^2=\frac{2}{3} \ .
\eea
It would be interesting to understand how this lower bound on $c$ gets modified at order $\alpha'$. Unfortunately, the $\rho$ potential becomes then too complicated to be able to obtain a simple inequality as the present one. We would then be back to include $\tau$ as well. It remains interesting to see that in the pure classical regime, we manage once again to saturate the TCC bound.\\

All our results are summarized in Table \ref{tab:c} and discussed in section \ref{sec:sumc}.

\section{Distance conjecture and $\lambda$ values}\label{sec:dist}

In this section we focus on the (refined) swampland distance conjecture introduced around \eqref{distconj}. We report on various examples in the literature where this conjecture has been tested, and we are particularly interested in the value of $\lambda$. We also compute $\lambda$ for few further examples and, after harmonising conventions on the distance definition, we precisely compare all of the $\lambda$ values obtained.

In order to test the distance conjecture, one first has to identify an infinite tower of states having the predicted properties. This is a non-trivial task, and there is often not only one tower of states becoming massless, but rather a competing collection of towers. Such states were studied e.g.~in \cite{Grimm:2018ohb, Blumenhagen:2018nts, Grimm:2018cpv, Gonzalo:2018guu, Corvilain:2018lgw, Joshi:2019nzi, Marchesano:2019ifh, Font:2019cxq, Erkinger:2019umg, Lee:2019wij, Enriquez-Rojo:2020pqm}. Further works \cite{Heidenreich:2016aqi, Heidenreich:2017sim, Andriolo:2018lvp, Heidenreich:2018kpg, Lee:2018urn, Lee:2018spm, Lee:2019tst, Lee:2019xtm} studied the relation of such towers to the weak gravity conjecture and the distance conjecture. Identifying a tower becoming massless is not enough: one further needs to show that the mass scale is suppressed exponentially in the distance, as given in \eqref{distconj}, and finally estimate $\lambda$. In the following, we will focus on works that obtained such results for Kaluza--Klein states \cite{Blumenhagen:2018nts, Erkinger:2019umg} or brane states \cite{Grimm:2018ohb, Font:2019cxq, Corvilain:2018lgw, Grimm:2018cpv, Joshi:2019nzi, Enriquez-Rojo:2020pqm}. This will provide us with estimates of $\lambda$ for different states and various (infinite distance) directions in the moduli space or field space.

To that end, let us first clarify the notion of distance ${\cal D}(P,Q)$ from a point $P$ to $Q$ on a (field) space, here a manifold. For real coordinates $x^i$, the metric can be expressed as $\d s^2 = g_{ij} \d x^i \d x^j $, and the distance along a curve $\gamma$ from $P$ to $Q$ parameterized by an affine parameter $s$ is given by
\beq
{\cal D}(P,Q) = \int_{\gamma} \sqrt{g_{ij} \frac{\del x^i}{\del s}\frac{\del x^j}{\del s}}\ \d s \ . \label{dist1}
\eeq
When the manifold is complex of real dimension $2r$, one can introduce complex holomorphic coordinates $z^{\alpha}$ and anti-holomorphic ones $\overline{z^\alpha}=\bar z^{\bar \alpha}$, $\alpha=1\dots r$, and the metric can then be rewritten as follows (see e.g.~appendix A.2 of \cite{Andriot:2011iw})
\beq
\d s^2 = g_{ij} \d x^i \d x^j  = 2  g_{\alpha \bar{\beta}} \d z^\alpha \d \bar z^{\bar{\beta}} + g_{\alpha \beta} \d z^\alpha \d z^\beta + g_{\bar{\alpha} \bar{\beta}} \d \bar z^{\bar{\alpha}} \d \bar z^{\bar{\beta}} \ .
\eeq
An Hermitian metric, to be considered in the following, is such that $g_{\alpha \beta}= g_{\bar{\alpha} \bar{\beta}}=0$, so the line element simplifies to $\d s^2 = 2  g_{\alpha \bar{\beta}} \d z^\alpha \d \bar z^{\bar{\beta}}$. The distance \eqref{dist1} then becomes
\beq
{\cal D}(P,Q) = \int_\gamma \sqrt{2\, g_{\alpha \bar{\beta}}\frac{\partial z^\alpha}{\partial s} \frac{\partial \bar z^{\bar{\beta}}}{\partial s}}\ \d s \ . \label{dist2}
\eeq
This definition of the distance, also compatible with the kinetic term convention of \eqref{Sddintro}, is the one to be used below. The factor $\sqrt{2}$ in this definition was not included in \cite{Blumenhagen:2018nts, Erkinger:2019umg, Joshi:2019nzi, Palti:2019pca}. We will consider in the following the values of $\lambda$ obtained in these works rescaled by this factor of $\sqrt{2}$.

\subsection{Kaluza--Klein states}

In \cite{Blumenhagen:2018nts, Erkinger:2019umg} concrete values of $\lambda$ were obtained by studying trajectories in the K\"ahler moduli space\footnote{To be precise in \cite{Blumenhagen:2018nts}  trajectories in the complex structure moduli space were studied and the results were then translated to the K\"ahler moduli space using mirror symmetry.} of Calabi--Yau (CY) hypersurfaces and complete intersections in projective spaces. The tower becoming massless was taken to be that of Kaluza--Klein states. As a result, the range of values of $\lambda$ obtained in these works (divided by $\sqrt{2}$ as explained above) is the following
\begin{equation}
0.6013 \leq \lambda \leq \sqrt{2} \ .
\end{equation}
The individual $\lambda$ values can be found in Table \ref{tab:lambda}. We indicate there various information on the references and the details of the model (space, phase, the number of K\"ahler moduli $h^{1,1}$). Some cases exhibit a range of values: this corresponds to different field space directions or different phases in which $\lambda$ has been computed; we refer to the original works for more details. Note also that in these references, $\Theta_\lambda = \frac{1}{\lambda}$ was given, while we give here the value of $\lambda$ directly (divided by $\sqrt{2}$). Finally, we take the opportunity to compute values for $\mathbb{P}^5[33],\mathbb{P}^5_{1^42^2}[44]$ and $\mathbb{P}^5_{1^22^23^2}[66]$, which to the best of our knowledge have not been calculated before. The results for $\mathbb{P}^4_{1^22^3}[8]$ and $\mathbb{P}^4_{1^22^26}[12]$ are very close, which is expected given that their moduli spaces are very similar.

In the 2 parameter models ($h^{1,1}=2$), it is more difficult to compute explicit geodesics. In \cite{Blumenhagen:2018nts}, the calculation was then done with an asymptotic expansion of the K\"ahler potential in the various phases, and the focus was on geodesics deep in a phase: see sections 5.1.3, 5.1.4 of \cite{Blumenhagen:2018nts} for more details. We simply list in Table \ref{tab:lambda} the results obtained in different phases.

\subsection{Brane states}

We now study the relation between brane states and the distance conjecture. We focus on the complex structure moduli space $\mathcal{M}_{cs}$ of type IIB string theory compactified on a Calabi-Yau threefold, CY${}_3$ (see e.g.~\cite{Candelas:1990pi} for a review). In general $\mathcal{M}_{cs}$ is not smooth and admits singularities. These special points can always be moved to lie at divisors which intersect normally \cite{heisuke, viehweg2012quasi}. In \cite{Grimm:2018ohb} singular points in $\mathcal{M}_{cs}$ were studied by means of the monodromy around these points.\footnote{\label{foot:dimGrimm}Part of the results of \cite{Grimm:2018ohb} also apply to CY of different complex dimension $D$; however some aspects of this work, such as the masses of the states, use ${\cal N}=2$ supersymmetry in $d=4$ dimensions, which from 10d type IIB requires to have $D=3$ . We thus restrict to these dimensions here.} In particular they used a theorem by Schmid \cite{Schmid1973} which allows one to get the asymptotic form of the K\"ahler potential near special points in $\mathcal{M}_{cs}$. In the one parameter case (one field) the expansion of the Weil--Petersson metric on the moduli space is given by
\beq
 g_{t \bar{t}} = \frac{1}{4} \frac{n}{(\operatorname{Im}t)^2} + \frac{\#}{(\operatorname{Im}t)^3}+ \dots + \mathcal{O}(e^{2 \pi \i t})\ , \label{gtt}
\eeq
where $n$, a nilpotency index,\footnote{Equivalently, $n$ is the degree of the polynomial $P({\rm Im}t)$ appearing in an expansion of the K\"ahler potential near a singular point \cite{Grimm:2018ohb}: $e^{-K}=P({\rm Im} t) + \mathcal{O}(e^{2 \pi \i t})$.} encodes the nature of the singular point and $t$ is the coordinate on $\mathcal{M}_{cs}$, such that the singular locus lies at $t\rightarrow \i \infty$. We refer to \cite{Grimm:2018ohb} and references therein for more details. Importantly, the leading term in \eqref{gtt} is universal, in the sense that it does not depend on the details of the moduli space. The distance displays now the following leading behaviour
\beq
{\cal D}(P,Q) = \int_{P}^{Q} \sqrt{2g_{t\bar{t}}}\ |\d t| \ \approx\ \sqrt{\frac{n}{2}}\, \ln \frac{\operatorname{Im} t|_Q}{\operatorname{Im} t|_P}\ \rightarrow\ \infty \ . \label{Dmcs}
\eeq
Furthermore, in \cite{Grimm:2018ohb} BPS states  were identified as possible candidate states which become massless, when approaching an infinite distance singularity. More precisely, the authors of \cite{Grimm:2018ohb} looked at BPS states in $\mathcal{N}=2$ supergravity originating from wrapped $D_3$-branes. The mass of these states is equal to the central charge, $M_{\mathbf{q}}=|Z_{\mathbf{q}}|$ with $\mathbf{q}$ the charge vector. From the comparison of the asymptotic form near singular points of the mass $M_{\mathbf{q}}$ of these BPS states, which become light, with the behaviour of the distance, an important result of \cite{Grimm:2018ohb} was
\beq
\frac{M_{\mathbf{q}}(Q)}{M_{\mathbf{q}}(P)} \simeq \frac{(\operatorname{Im}t)^s|_P}{(\operatorname{Im}t)^s|_Q} \simeq e^{-\lambda\ {\cal D}(P,Q)}\ ,
\eeq
where the second equality is obtained from \eqref{Dmcs}, with
\beq
\lambda = s\sqrt{\frac{2}{n}} \quad s=\begin{cases}1 & \text{if} \quad n \mod 2 = 0 \\
       \frac{1}{2} & \text{if} \quad n \mod 2 \neq 0\end{cases} \ .
\eeq
Note that $n$ is bounded by the complex dimension of the CY, i.e.~here $n=\{1,2,3\}$, giving the following three possible values
\beq
\lambda =\left\{ \frac{1}{2} \sqrt{2} \ ,\ \ 1\ ,\ \ \frac{1}{2} \sqrt{\frac{2}{3}} \right\}\ ,
\eeq
that we report in Table \ref{tab:lambda}.

These remarkable general results were reproduced in various concrete examples.\footnote{Our short review of the results of \cite{Grimm:2018ohb} corrects some typos present in that paper; in addition, the results of \cite{Joshi:2019nzi} are here rescaled with the $\sqrt{2}$ factor mentioned at the beginning of this section.} For one parameter models as here, it is sufficient to give the $\lambda$ values for one class of examples, namely the complete intersection cases of the form $\mathbb{P}^5 [cc]$ with two infinite distance points, because all remaining models have the same type of singularity at the large complex structure point ($LCS$). The  small complex structure ($SCS$) point is at finite distance for the remaining models. All one parameter models were actually studied in detail in \cite{Joshi:2019nzi}, where the following results were obtained
\begin{align}
\mathbb{P}^5 [cc]: &  & \lambda_{LCS} &=\frac{1}{\sqrt{6}} \ , & \lambda_{SCS} &= \frac{1}{\sqrt{2}}\ ,
\end{align}
for $D_3$-branes at the $SCS$ point in the $\mathbb{P}^5[33]$ model and mirror $D_0$-$D_2$ bound states at the $LCS$ point. In the latter, $D_0$-branes are less dominant with $\lambda=\sqrt{3/2}$.

Various extensions were considered in \cite{Grimm:2018ohb}. In general, having determined possible states, one still has to show that there is an infinite tower of them and verify if they are stable against decay. A systematic study of the existence of such towers at different types of singularities has been performed in \cite{Grimm:2018ohb, Grimm:2018cpv}. Similar questions have been addressed from the perspective of the K\"ahler moduli space in \cite{Corvilain:2018lgw}. Finally, an analogous analysis can be found in \cite{Font:2019cxq, Enriquez-Rojo:2020pqm} where possible states were studied in type IIA/IIB orientifold compactifications, i.e.~with ${\cal N}=1$ supersymmetry in 4d. In particular, considering $O_3/O_7$ on CY${}_3$ in type IIB \cite{Enriquez-Rojo:2020pqm}, the values $\lambda=1/\sqrt{6}$ and $\lambda=\sqrt{3/2}$ were also obtained with $D_3$-branes. We summarize all these results in Table \ref{tab:lambda}.

\subsection{Comments and $\lambda$ values}

All $\lambda$ values for the distance conjecture \eqref{distconj} obtained in the various settings discussed above are now summarized in Table \ref{tab:lambda}. Let us give few comments on these values. A first striking point is that all values verify
\beq
\lambda\ \geq\ \frac{1}{2}\, \sqrt{\frac{2}{3}}\ \approx\ \frac{1}{2}\, 0.8165\ \approx\ 0.4082 \ , \label{lambdabound}
\eeq
with several examples of saturation. A very convincing case is the general analysis of \cite{Grimm:2018ohb} on CY${}_3$ in type IIB with $n=3$, that provides precisely this value. The result \eqref{lambdabound} is remarkable because it corresponds exactly to a half of the TCC bound \eqref{TCCbound}, verified by $c$ in the de Sitter no-go theorems as discussed in section \ref{sec:sumc}. While it was already surprising for these no-go theorems, it is here at first sight extremely puzzling for the distance conjecture. It is fair to say that studies of CY${}_3$ moduli space and their BPS states seem a priori perfectly unrelated to the TCC concepts and framework, and the numbers obtained in \cite{Grimm:2018ohb} could have been completely different, even if close to $1$. The swampland perspective may however provide some clarification, as we will discuss in section \ref{sec:web}.

Let us make few more comments on the $\lambda$ values in Table \ref{tab:lambda}. It is interesting to note that twice the bound \eqref{lambdabound} is also sometimes obtained, and similarly with the value $\sqrt{2}$. The same values and factor of 2 were observed as well for $c$, as can be seen in Table \ref{tab:c}. The matching of these $c$ and $\lambda$ values will be further discussed in section \ref{sec:web}. Finally, regarding the numerical values appearing in the ranges of $\lambda$, it is interesting to note that the lowest value of the upper range bounds is $0.8165$, which is precisely the estimate of $\sqrt{\frac{2}{3}}$. Other upper bounds are close to $1$, which is another analytical value which appears. We are however not sure how to interpret these points nor the lower range bounds; it remains remarkable that $\sqrt{\frac{2}{3}}$ appears there once again.

\begin{table}[H]
  \begin{center}
    \begin{tabular}{|c|c|c||c|}
    \hline
    Tower states & Setting & Reference & $\lambda$  \\
    \hhline{====}
     & & &  \\[-8pt]
     & $\mathbb{P}^4[5]$, $h^{1,1}=1$ & Table 3 of \cite{Blumenhagen:2018nts} & $0.7168 \leq \lambda \leq 0.8168$ \\[3pt]
    \hhline{~--||-}
     & & & \\[-8pt]
     & $\mathbb{P}^4_{1^42}[6]$, $h^{1,1}=1$ & Table 4 of \cite{Blumenhagen:2018nts} & $0.7389 \leq \lambda \leq 0.8165$ \\[3pt]
    \hhline{~--||-}
     & & & \\[-8pt]
     & $\mathbb{P}^4_{1^44}[8]$, $h^{1,1}=1$ & Table 5 of \cite{Blumenhagen:2018nts} & $0.7579 \leq \lambda \leq	0.8175$ \\[3pt]
    \hhline{~--||-}
     & & & \\[-8pt]
     & $\mathbb{P}^4_{1^325}[10]$, $h^{1,1}=1$ & Table 6 of \cite{Blumenhagen:2018nts} & $0.7451 \leq \lambda \leq 0.8175$ \\[3pt]
    \hhline{~--||-}
     & & & \\[-8pt]
     & $\mathbb{P}^6_{1^7}[322]$, $h^{1,1}=1$ & Table 4 of \cite{Erkinger:2019umg} & $0.6013 \leq \lambda \leq 1.0177$ \\[3pt]
    \hhline{~--||-}
     & & & \\[-8pt]
     & $\mathbb{P}^5_{1^6}[42]$, $h^{1,1}=1$ & Table 5 of \cite{Erkinger:2019umg} & $0.6121 \leq \lambda \leq 0.9650$ \\[3pt]
    \hhline{~--||-}
     & & & \\[-8pt]
     & $\mathbb{P}^5_{1^53}[62]$, $h^{1,1}=1$ & Table 6 of \cite{Erkinger:2019umg} & $0.6654 \leq \lambda \leq 1.0425$ \\[3pt]
    \hhline{~--||-}
     & & & \\[-8pt]
    Kaluza--Klein & $\mathbb{P}^5_{1^52}[43]$, $h^{1,1}=1$  & Table 7 of \cite{Erkinger:2019umg} & $0.6591 \leq \lambda \leq 0.9600$ \\[3pt]
    \hhline{~--||-}
    states & & & \\[-8pt]
     & $\mathbb{P}^5[33]$, $h^{1,1}=1$ & new & $0.7834 \leq \lambda \leq 0.8837$ \\[3pt]
    \hhline{~--||-}
     & & & \\[-8pt]
     & $\mathbb{P}^5_{1^42^2}[44]$, $h^{1,1}=1$ & new & $0.7835 \leq \lambda \leq 0.9182$ \\[3pt]
    \hhline{~--||-}
     & & & \\[-8pt]
     & $\mathbb{P}^5_{1^22^23^2}[66]$, $h^{1,1}=1$ & new & $0.7828 \leq \lambda \leq 0.9447$ \\[3pt]
    \hhline{~--||-}
     & & & \\[-10pt]
     & $\mathbb{P}^4_{1^22^3}[8]$, $h^{1,1}=2$, Hybrid-Orbifold & Section 5.1.4 of \cite{Blumenhagen:2018nts} & $\sqrt{\frac{2}{3}}$ \\[5pt]
    \hhline{~--||-}
     & & & \\[-8pt]
     & $\mathbb{P}^4_{1^22^3}[8]$, $h^{1,1}=2$, Hybrid-$\mathbb{P}^1$ & Section 5.1.4 of \cite{Blumenhagen:2018nts} & $\sqrt{2}$ \\[3pt]
    \hhline{~--||-}
     & & & \\[-10pt]
     & $\mathbb{P}^4_{1^22^26}[12]$, $h^{1,1}=2$, Hybrid-Orbifold & Section 5.2 of \cite{Blumenhagen:2018nts} & $\sqrt{\frac{2}{3}}$ \\[5pt]
    \hhline{~--||-}
     & & & \\[-8pt]
     & $\mathbb{P}^4_{1^22^26}[12]$, $h^{1,1}=2$, Hybrid-$\mathbb{P}^1$ & Section 5.2 of \cite{Blumenhagen:2018nts} & $\sqrt{2}$ \\[3pt]
    \hhline{~--||-}
     & & & \\[-10pt]
     & $\mathbb{P}^4_{1^369}[18]$, $h^{1,1}=2$, Hybrid-Orbifold & Section 5.3 of \cite{Blumenhagen:2018nts} & $\sqrt{\frac{2}{3}}$ \\[5pt]
    \hhline{~--||-}
     & & & \\[-8pt]
     & $\mathbb{P}^4_{1^369}[18]$, $h^{1,1}=2$, Hybrid-$\mathbb{P}^2$ & Section 5.3 of \cite{Blumenhagen:2018nts} & $1$ \\[3pt]
    \hhline{---||-}
    & & & \\[-8pt]
     & CY${}_3$, type IIB, $n=1$ & \cite{Grimm:2018ohb} & $\frac{1}{2} \sqrt{2}$ \\[3pt]
    \hhline{~--||-}
    & & & \\[-8pt]
     & CY${}_3$, type IIB, $n=2$ & \cite{Grimm:2018ohb} & $1$ \\[3pt]
    \hhline{~--||-}
    & & & \\[-10pt]
    Brane & CY${}_3$, type IIB, $n=3$ & \cite{Grimm:2018ohb} & $\frac{1}{2} \sqrt{\frac{2}{3}}$ \\[5pt]
    \hhline{~--||-}
    states & & & \\[-10pt]
     & $\mathbb{P}^5 [cc]$, large complex struct. point & (6.5) of \cite{Joshi:2019nzi} & $\frac{1}{2} \sqrt{\frac{2}{3}}$ \\[5pt]
    \hhline{~--||-}
    & & & \\[-8pt]
     & $\mathbb{P}^5 [cc]$, small complex struct. point & (6.22) of \cite{Joshi:2019nzi} & $\frac{1}{2} \sqrt{2}$ \\[3pt]
    \hhline{~--||-}
    & & & \\[-10pt]
     & CY${}_3$ with orientifold, type IIB & Section 4.2 of \cite{Enriquez-Rojo:2020pqm} & $\frac{1}{2} \sqrt{6}$, $\frac{1}{2} \sqrt{\frac{2}{3}}$ \\[5pt]
    \hline
       \end{tabular}
     \caption{Values of $\lambda$ in the distance conjecture \eqref{distconj} that were obtained in the literature, as well as few new ones. The states of the tower becoming massless are either Kaluza--Klein states or brane states. For each value or range of values, we give the setting in which it was obtained and the reference in the literature. More details can be found in the main text.}\label{tab:lambda}
  \end{center}
\end{table}

\section{The web of conjectures}\label{sec:web}

We have observed in previous sections that the two numbers of order one in the de Sitter conjecture and in the distance conjecture, namely $c$ and $\lambda$, both admit a lower bound, at least in the examples analysed. For $c$, this bound is consistent with the TCC bound \eqref{TCCbound}, but this result remains impressive. For $\lambda$, to our surprise, it matches a half of the TCC bound \eqref{lambdabound}. In addition, further values obtained for these numbers, listed in Table \ref{tab:c} and \ref{tab:lambda}, also match up to a $\frac{1}{2}$ factor. In our view, these results can hint at a relation between the two conjectures, that we will present as a map and discuss in detail. The idea of such a relation is not new, and we will first review in section \ref{sec:weblit} similar ideas already present in the literature. The lower bound on $\lambda$ also calls, in our view, for a modification of the distance conjecture that we will detail. All of our proposals on swampland conjectures and their relations are then given in section \ref{sec:prop}; that section will necessarily be more speculative than the rest of this paper. We finally discuss important aspects of the aforementioned map and provide non-trivial checks in section \ref{sec:checkmap}. In the entire section, we work again in Planckian units.

\subsection{Relating conjectures in the literature}\label{sec:weblit}

It has been argued and tested that the swampland conjectures are not independent of each other but form a web \cite{Palti:2019pca}. Relations between conjectures are not all established on the same footing, some are stronger than others. As we will see, the de Sitter swampland conjecture(s) have been argued to be related to some forms of the distance conjecture or the weak gravity conjecture, even though not in a very precise manner. Given the results of the previous sections, we are interested in making such a relation more precise. To this end, we first give here a brief overview of the literature on this matter.

Let us start with some clarifications. On the one hand, the distance conjecture is a statement that was mostly checked on Calabi--Yau compactifications (typically to 4d Minkowski) and their moduli spaces. On the other hand, the de Sitter statement is typically verified in compactifications on curved manifolds with fluxes and sources, generating a non-trivial 4d scalar potential $V$ and a cosmological constant $\Lambda$. The two setups are therefore at first sight very different. The conjectures are however supposed to hold in any setup. It is then unclear whether a relation between the two statements could be established in any framework, or even in a single setup. Rather, one possibility is to view the relation between the conjectures as a relation between two different setups or theories, which can be understood to some extent as a duality. This idea was sketched in \cite{Ooguri:2018wrx}. If one was to consider both conjectures in a single setup, it would be tempting to compare the mass entering the distance conjecture, $m(\varphi)$, to the mass of $\varphi$ itself, related to $\del^2_{\varphi} V$ for a scalar potential $V(\varphi)$. These two masses are however those of different states, thus they have a priori no relation. Rather we think on general grounds that the former $m$ could be compared to $V$ itself (in Planckian units), and we will further elaborate on this.\\

One guideline followed in the literature when comparing conjectures are the exponentials that appear as saturating cases of certain inequalities. For example, in the present work we have already found them in (the saturation of) \eqref{V'/V}, \eqref{TCC} and \eqref{distconj}. We are especially interested in a relation between the $\mathcal{O}(1)$ numbers in these exponentials. Since these exponentials appear in large field limits, those limits should also play a role in the relation between conjectures. It was argued in \cite{Ooguri:2018wrx} that such limits typically correspond to parametrically controlled regimes in string theory where, considering the whole string framework, one has at his disposal a dual description. It could then be argued that one conjecture would apply on one side of the duality, and could be mapped (by identifying the exponential behaviour of some quantity) to another side into a different statement. In \cite{Ooguri:2018wrx}, such an idea was used to count the number of light states (see e.g.~\cite{Corvilain:2018lgw}) contributing to the entropy of a quasi-de Sitter space-time. In the large field limit, this led to the identification of the asymptotic form of the scalar potential $V$ with the exponential behaviour of $m$ (modulo a function counting the number of towers becoming light). In this relation, the power of the exponential was however not given explicitly and seemed dependent on the context. From these ideas, we still infer the possibility of having the following type of relation
\beq
\frac{m}{m_i} \approx \left|\frac{V}{V_i}\right|^{\alpha} \approx e^{- \lambda \, {\cal D}}  \quad \mbox{when}\ {\cal D} \rightarrow \infty  \ , \label{mVrel}
\eeq
with some constants $m_i, V_i, \alpha$ and $\mathcal{D}$ the geodesic field distance. Considering the saturation case in the de Sitter conjecture, we deduce a possible equality
\beq
\lambda = \alpha\, c  \ .
\eeq

Another swampland conjecture of interest is the scalar weak gravity conjecture \cite{Palti:2017elp}. Considering a field of mass $m$ coupled to a scalar field $\varphi$ through a Yukawa coupling, the conjecture requires gravity to be the weakest force, in the spirit of the original weak gravity conjecture \cite{ArkaniHamed:2006dz}. This is to be distinguished from the requirement of avoiding bound states, which implies that repulsive forces overcome attractive ones \cite{Lee:2018spm, Heidenreich:2019zkl}, and leads to different modifications of the weak gravity conjecture due to scalars. Comparing the Yukawa coupling to that of gravity, the scalar weak gravity conjecture is written as follows for a single scalar field $\varphi$
\beq
(\del_{\varphi} m)^2 \geq m^2 \ . \label{swgc}
\eeq
Interestingly, this inequality is very similar in form to the de Sitter conjecture \eqref{V'/V}; we will come back to this point. The saturation of the inequality also gives back the exponential of the distance conjecture. This will be another important relation in the following. A generalization of this conjecture was proposed in \cite{Gonzalo:2019gjp} under the name ``strong scalar weak gravity conjecture'' (see \cite{DallAgata:2020ino} for a recent extension). This new conjecture applies such an inequality to self-interactions of a single scalar field $\varphi$ governed by a potential $V(\varphi)$. It follows from the idea that $m^2$ in \eqref{swgc} could be traded for the second derivative $V''$, and the inequality would get corrected by quartic interactions. The result would be
\beq
2 \frac{(V''')^2}{V''} - V'''' \geq V'' \quad \leftrightarrow\quad  (V'')^2 \left(\frac{1}{V''}\right)'' \geq V''  \ .\label{strongswgc}
\eeq
It is again interesting to consider the saturation case, where one obtains
\beq
\left(\frac{1}{V''}\right)'' - \frac{1}{V''} = 0 \ \leftrightarrow \ V'' = \left(A e^{\varphi} + B e^{-\varphi}  \right)^{-1} \ ,
\eeq
with two integration constants $A,B$.\footnote{One may generalize \eqref{strongswgc} and its saturation to allow for different coefficients, namely
\beq
n \frac{(V''')^2}{V''} - V'''' = V''  \leftrightarrow  \frac{1}{n-1} \left(\frac{1}{(V'')^{n-1}}\right)'' - \frac{1}{(V'')^{n-1}} = 0  \ \mbox{for}\ n\in \mathbb{R},\ n \neq 1 \ .\nn
\eeq} Therefore, in the saturation case one recovers the exponentials.\footnote{Interestingly, these two exponentials are mapped into one another by $e^\varphi\to(e^\varphi)^{-1}$, a fact which is reminiscent of the relation between momentum and winding modes of the string under T-duality.} Trading $V''$ for $m^2$, one obtains again a distance conjecture behaviour. Rewriting \eqref{strongswgc} as
\beq
m^2 \left(\frac{1}{m^2}\right)'' \geq 1 \ , \label{mIb}
\eeq
it has been further proposed in the unpublished work \cite{TalkIbanez} to trade $m$ for the scalar potential $V$ of a de Sitter conjecture. More precisely, a relation $V = m^{2\gamma}$ for some unspecified number $\gamma$ was proposed, which is nothing but the first relation in \eqref{mVrel}. Doing so, the inequality \eqref{mIb} translates into a refined de Sitter conjecture \`a la \cite{Andriot:2018wzk}, when allowing in \cite{Andriot:2018wzk} the limiting case $q=2$. This web of conjectures is interesting, and illustrates the various key ideas in this discussion. A crucial role is played by an identification of the type \eqref{mVrel} between $m$ and $V$, in order to connect the distance conjecture (and related conjectures) to a de Sitter conjecture.

Finally, a conjecture has been recently proposed regarding the distance conjecture in an anti-de Sitter space-time \cite{Lust:2019zwm} (see also \cite{Gautason:2018gln} and the recent \cite{Font:2019uva, Buratti:2020kda}). It states that in the near-flat limit, where the cosmological constant $\Lambda \rightarrow 0$, a tower of states of mass scale $m$ becomes light, obeying
\beq
m \sim |\Lambda|^{\alpha} \quad \mbox{for}\ \Lambda \rightarrow 0 \ . \label{ADC}
\eeq
It is further suggested that $\alpha \geq  \frac{1}{2}$, and one should have $\alpha = \frac{1}{2}$ for supersymmetric vacua (strong version). Some discussion in \cite{Lust:2019zwm} on an extension to de Sitter allows various options for $\alpha$, including the value $\frac{1}{2}$ as a bound. Here, by extrapolating this conjecture to quasi-de Sitter or anti-de Sitter space-times, we replace $\Lambda$ by $V$ in \eqref{ADC} and then reproduce, up to extra constant factors, the relation \eqref{mVrel} in an asymptotic limit. This can motivate once more the tentative \eqref{mVrel}. A difference though is that the tower of states and the cosmological constant $\Lambda$ are considered in \cite{Lust:2019zwm} in the same setup, while here we argued at first for such a map between two different setups.

Further works proposing interesting relations between all these conjectures include \cite{Brahma:2019mdd, Brahma:2019vpl, Geng:2019zsx}. Having reviewed ideas on the web of conjectures discussed in the literature, we now focus on our results.

\subsection{Proposals}\label{sec:prop}

Inspired by the ideas reviewed above, as well as our results discussed around Table \ref{tab:c} and \ref{tab:lambda}, we now propose several conjectured statements.

\subsubsection{Distance conjecture}\label{sec:propdist}

We first propose to relax the distance conjecture \eqref{distconj} to a weaker, thus more general, statement. It takes a TCC form \eqref{TCC} and is inspired by the bound on $\lambda$ \eqref{lambdabound} as well as by a map to the potential $V$ of the de Sitter conjecture. We propose the following statement\\

{\it Consider a field space (e.g.~a scalar moduli space) appearing in a $d$-dimensional low energy effective theory of a quantum gravity, and a geodesic distance ${\cal D}$ in this field space. Whenever ${\cal D} \rightarrow \infty$, a tower of states becomes light, with a typical mass scale $m$ verifying in Planckian units}
\beq
 0 < m \leq m_0\ e^{- \lambda_0 \, {\cal D}} \ ,\label{distconjus}
\eeq
{\it with positive constants $m_0,\ \lambda_0$, where}
\beq
\lambda_0 = \frac{1}{\sqrt{(d-1) (d-2)}} \ . \label{lambda0value}
\eeq\\

Consider that the distance is given as a function of a single canonically normalized field $\varphi$, namely ${\cal D}=|\varphi - \varphi_i|$. Taking without loss of generality $\varphi > \varphi_i$, one has
\beq
-\left\langle \frac{m'}{m} \right\rangle \, =\, -\frac{1}{\varphi - \varphi_i} \int_{\varphi_i}^{\varphi} \d \tilde{\varphi}\, \frac{\del_{\tilde{\varphi}} m}{m} \, =\,  -\frac{\ln \frac{m(\varphi)}{m(\varphi_i)}}{\varphi - \varphi_i} \, \geq \, -\frac{\ln \frac{m_0}{m(\varphi_i)}}{\varphi - \varphi_i} + \lambda_0 \ , \label{computaverage}
\eeq
where in the last step we used \eqref{distconjus}. We further observe, by definition and using the formula above, that
\beq
\left\langle \frac{|m'|}{m} \right\rangle \, \geq \, \left|\left\langle \frac{m'}{m} \right\rangle\right| \, \geq \, \frac{|\ln \frac{m_0}{m(\varphi_i)}|}{{\cal D}} + \lambda_0 \ ,
\eeq
for a large distance. We conclude that
\beq
\eqref{distconjus} \ \Rightarrow\ \left\langle \frac{|m'|}{m} \right\rangle_{{\cal D} \rightarrow \infty}\, \geq\, \lambda_0 \ . \label{distconjus2}
\eeq
This derivation mimics that of the TCC \eqref{TCC} in \cite{Bedroya:2019snp}, without assuming any sign for $m'$. As in \cite{Bedroya:2019snp}, the first derivative $m'=\del_{\varphi} m$ gets naturally replaced in a multi-field case by $\nabla m$. The distance is then defined along a geodesic path in the (multi-)field space. Importantly, the inequality \eqref{distconjus2} is nothing but a (generalized) scalar weak gravity conjecture \eqref{swgc}. We will come back to this point below.

The inequality \eqref{distconjus} includes as a particular case the situation when the standard distance conjecture \eqref{distconj} is verified,\footnote{An inequality which reminds \eqref{distconjus}, though it has some differences, actually appears in \cite{Palti:2019pca}, where the refined distance conjecture is presented. However, we could not find that inequality explicitly in the original papers \cite{Klaewer:2016kiy, Baume:2016psm}, so its motivation is unclear to us. The main difference with \eqref{distconjus} is that the exponent in \cite{Palti:2019pca} is a generic number of order one, while in our case the bounding exponential is fixed, with a definite constant $\lambda_0$. This and the fact that the review \cite{Palti:2019pca} appeared before the TCC paper \cite{Bedroya:2019snp}, are to us important conceptual differences between these inequalities. Finally, a similar inequality is also sketched in \cite{Bedroya:2019snp}, with however a different exponent.} namely when the mass scale goes like an exponential in the large distance limit ${\cal D} \rightarrow \infty$. In that case, we deduce from \eqref{distconjus} or \eqref{distconjus2} the following constraint
\beq
m = m_i\ e^{- \lambda \, {\cal D}} \quad \Rightarrow \quad \lambda\, \geq\, \lambda_0 \ . \label{ll0}
\eeq
Another way to understand the proposal \eqref{distconjus} is that it allows $m$ to have an exponential behaviour in various directions, but these exponentials should be subdominant compared to the saturation case of the inequality. In particular, one deduces the bound
\beq
\lambda \, \geq\, \frac{1}{2}\, \sqrt{\frac{2}{3}}  \quad \mbox{in}\ d=4 \ , \label{lambdabound2}
\eeq
which was verified very precisely in all examples summarized in Table \ref{tab:lambda}, as discussed around \eqref{lambdabound}. In the framework of \cite{Grimm:2018ohb}, this bound was actually proven to hold in general for wrapped $D_3$-brane states. This motivates in $d=4$ the value of $\lambda_0$ \eqref{lambda0value} being a half of the TCC one \eqref{TCC}, i.e.~$\lambda_0 = \frac{1}{2} c_0$. The extension to other $d$ seems natural, and further motivated by the proposal of section \ref{sec:proprel}, even though we did not check it here against concrete examples. As detailed in footnote \ref{foot:dimGrimm}, we did not find  for $d\neq4$ a counter-example to this value in the distance conjecture literature.

The inequality \eqref{distconjus} also allows for different behaviours of the mass $m$ than an exponential. The examples considered in this paper present only exponentials, but the idea of allowing logarithmic corrections was recently put forward following \cite{Bedroya:2019snp}, especially for masses in e.g.~\cite{Blumenhagen:2019vgj}. Indeed, one already sees in the ratio $m'/m$  in \eqref{computaverage} the appearance of some logarithmic corrections, that disappear in the limit of infinite distance in \eqref{distconjus2}. The extra room allowed by the inequality \eqref{distconjus}, compared to the distance conjecture \eqref{distconj}, is then interesting, e.g.~for what concerns quantum corrections. However, while the TCC inequality \eqref{TCC} on the potential $V$ was originally derived from a physical argument, we are lacking for now of an analogous argument behind \eqref{distconjus}; we hope to come back to this question in future work.

\subsubsection{Scalar weak gravity conjecture}\label{sec:propswgc}

Our generalization \eqref{distconjus} of the distance conjecture implies the inequality \eqref{distconjus2} that we repeat here for convenience
\beq
\left\langle \frac{|\del_{\varphi} m|}{m} \right\rangle_{{\cal D} \rightarrow \infty}\, \geq\, \lambda_0 \ \label{distconjus3}
\eeq
and that has the form of a scalar weak gravity conjecture. We thus promote the scalar weak gravity conjecture inequality \eqref{swgc} to the more general one \eqref{distconjus3}. Its relation to the distance conjecture, reviewed before, is now even more apparent. As discussed, the inequality \eqref{distconjus3} allows for more general behaviours of $m$. In particular cases where neither the average nor the limit is needed, e.g.~for $m$ being an exponential, the scalar weak gravity conjecture inequality \eqref{distconjus3} boils down to
\beq
(\del_{\varphi} m)^2 \geq \lambda_0^2\ m^2 \ , \label{swgcmodif}
\eeq
with a straightforward multi-field generalization by replacing the left-hand side with $(\nabla m)^2$. With respect to the original scalar weak gravity conjecture \eqref{swgc}, the novelty is now only the parameter $\lambda_0$. Given that in 4d, one has $1 > \lambda_0$, this modified version \eqref{swgcmodif} is then verified by all cases where the original conjecture \eqref{swgc} holds. Our modification suggests the possible existence of other cases to be discovered where this lower bound $\lambda_0$ will be reached. As noticed already for the distance conjecture, the value obtained in these inequalities depends on the compactification setup and on the field direction.

Finally, it would be natural to modify known extensions of the scalar weak gravity conjecture in the same manner. For example, one could consider the proposal \eqref{mIb} of \cite{Gonzalo:2019gjp}, and turn it into
\beq
m^2 \left(\frac{1}{m^2}\right)'' \geq \lambda_0^2 \ .
\eeq
It would be more satisfactory to ``derive'' such a modification analogously to what was done for \eqref{distconjus3} from \eqref{distconjus}. To this end, studying $(m^2)''$ could be useful, as it would involve quartic couplings. This is reminiscent of the study of $V''$ in \cite{Andriot:2018mav} that led to a refined de Sitter conjecture.

\subsubsection{Relating the distance and de Sitter conjectures}\label{sec:proprel}

We have reviewed in section \ref{sec:weblit} possible relations between the distance and the de Sitter conjectures, with an emphasis on a link between $m$ and $V$. If one allows for the inequality version of the distance conjecture \eqref{distconjus}, and given the TCC inequality \eqref{TCC}, the analogy between $m$ and $V$ becomes striking. The relations between the values of $c$ in Table \ref{tab:c} and $\lambda$ in Table \ref{tab:lambda}, as well as their lower TCC bounds, are to us an additional hint. The values of $c$ and $\lambda$ are very dependent on each side on details of the compactification and the chosen field direction. In addition, as discussed at the beginning of section \ref{sec:weblit}, the two conjectures seem to operate in different compactification setups, at least in the examples analysed. Having all these ideas in mind, we propose the following correspondence or map.\\

{\it We conjecture the existence of a map between two compactification setups to $d$ dimensions where, in each of them, a direction $\varphi_k$ in field space is selected: (setup${}_1$, $\varphi_1$) $\leftrightarrow$ (setup${}_2$, $\varphi_2$). In the first setup, the generalized distance conjecture \eqref{distconjus} with mass $m$ applies, and in the second one the de Sitter conjecture in TCC form \eqref{TCC} applies. We denote generically by ${\cal D}$ the field space geodesic distance along $\varphi_{k=1,2}$; it can on both sides be arbitrarily large. The proposed map is then}
\beq
\frac{m}{m_i} \simeq \left|\frac{V}{V_i}\right|^{\frac{1}{2}} \quad \mbox{for}\ {\cal D} \rightarrow \infty \ , \label{ourconj}
\eeq
{\it for some constants $m_i, V_i$. The symbol $\simeq$ is understood as an equality of the two functions $m, V$ up to the exchange $\varphi_1 \leftrightarrow \varphi_2$.}\\

When the mass and the potential are exponentials in the large field distance, both sides of \eqref{ourconj} are equal to $e^{- \lambda \, {\cal D}}$. We then match the tentative relation \eqref{mVrel} with the value $\alpha = \frac{1}{2}$ and  we also have
\beq
\lambda = \frac{1}{2} \, c\ \geq\ \lambda_0 = \frac{1}{2} \, c_0 \ , \label{relclambda}
\eeq
We will comment further on the number $c$ and on the asymptotic (exponential) form of $V$ in section \ref{sec:checkmap}. This relation between $\lambda$ and $c$ is well illustrated by pairs of examples in Table \ref{tab:c} and \ref{tab:lambda}. More generally, it is straightforward to infer from \eqref{ourconj} and for a single field that $|m'|/m \simeq \frac{1}{2}\, |V'|/V$ in the large distance limit. This relation generalizes \eqref{relclambda} beyond the exponential case. One could also define $\lambda$ and $c$ as the minimum of these ratios (or their average) in the large distance limit, consistently with the bounds discussed previously.

Note that the exponent $\frac{1}{2}$ in \eqref{ourconj} is consistent with the (strong) anti-de Sitter distance conjecture \cite{Lust:2019zwm} and extensions discussed around \eqref{ADC}. In the latter though, the setups related by the above map would actually be the same, while we allow here in general for two different ones. In addition, we divide here by constants, while it is not the case in \eqref{ADC}. As a consequence, we have in general nothing to say about scale separation. The anti-de Sitter distance conjecture can still be viewed as a particular case of \eqref{ourconj}, where the scales $m_i$ and $|V_i|^{\frac{1}{2}}$ are precisely the same and the two setups are identified. When the scales are different, the resulting multiplicative constant may account for the one discussed in \cite{Font:2019uva}. However, we recall that, strictly speaking, we can only discuss a de Sitter version of the anti-de Sitter distance conjecture, which was only sketched in \cite{Lust:2019zwm} and mentioned around \eqref{ADC}. Indeed, our map assumes a positive potential through the TCC \eqref{TCC}. Nevertheless, we find this connection to the anti-de Sitter distance conjecture satisfactory, to say the least.

We provide additional detailed checks and discussions on this map in the section \ref{sec:checkmap}. However, a more fundamental reason underlying the existence of this map remains unclear to us at this stage. We also comment further on this in the next section.

\subsection{Asymptotic exponential form of $V$, the number $c$ and more on the map}\label{sec:checkmap}

In section \ref{sec:dSsection}, we presented no-go theorems in the form of inequalities of the type \eqref{combinogo}, involving a linear combination of fields. This linear combination can be viewed as a single field direction. For future purposes, we find it convenient to denote it by $\varphi_2$, and rewrite the generic no-go inequality \eqref{combinogo} as follows
\beq
a\, V + \sum_{i} b_i\, \del_{\hat{\varphi}^i} V \leq 0 \quad \leftrightarrow\quad  c\, V + \del_{\varphi_2} V \leq 0 \ .\label{combinogophi2}
\eeq
Then, the technique to get a value for $c$ in section \ref{sec:dSsection} essentially amounts to use the following inequalities (considering $V>0$)
\beq
\frac{|\nabla V|}{V} \geq \frac{|\del_{\varphi_2} V|}{V} \geq c  \ ,
\eeq
and, in section \ref{sec:dSsection}, we further verified that $c \geq c_0$. This discussion highlights the fact that $c$, as defined above, is dependent on the selected field direction. This is somehow different from the original de Sitter conjecture \eqref{V'/V}, which rather considered $c$ as a universal bound, a role played in our discussion by $c_0$.\footnote{The number $c$ is here not equal to the minimum value of $|\nabla V|/V$, but rather to the minimum of $|\del_{\varphi_2} V|/V$ along a direction $\varphi_2$. The latter is a priori different and smaller than the former.} The Table \ref{tab:c} emphasizes already that $c$ depends on the compactification setup, i.e.~the no-go assumptions, but these are intimately related to the no-go inequality \eqref{combinogophi2} and thus to a field direction.

This situation is very analogous to that of the distance conjecture. There as well, the behaviour of the mass, especially for what concerns the value of $\lambda$, depends on the field direction in the moduli space. The nature of the tower having this mass also depends on the field direction. This is highlighted by the different lines in Table \ref{tab:lambda}, which in turn indicate the dependence of the field space on the compactification setup. Therefore, when trading $\lambda$ for $c$, one should exchange also the associated field directions. This point is further enforced by the map \eqref{ourconj} as we now detail.

In all examples on the distance conjecture considered in this paper and summarized in Table \ref{tab:lambda}, the mass $m$ verifies the exponential behaviour \eqref{distconj}.  In that case, the map \eqref{ourconj} implies that also the associated potential $V$ has to have an exponential  behaviour in the large distance limit. More precisely, we should have that
\beq
V = V_i\ e^{- c \, {\cal D}} \ \mbox{for}\ {\cal D} \rightarrow \infty \ .\label{Vcheck}
\eeq
where ${\cal D}=|\varphi_2 - \varphi_i|$,  $V_i, \varphi_i$ are some constants, and $\varphi_2$ is the field (direction) mentioned in the map \eqref{ourconj}. The fact that the exponent is precisely $c$ was already announced in \eqref{relclambda}, and is motivated by \eqref{combinogophi2}. Indeed, in the case of an exponential, the right-hand side inequality of \eqref{combinogophi2} calls for a saturation, fixing the exponent to be $c$. We now want to check the behaviour \eqref{Vcheck} on some examples, and thus verify the map \eqref{ourconj}. In this respect, a non-trivial step is to identify the appropriate field direction $\varphi_2$ to be sent to infinity. As anticipated above with \eqref{combinogophi2}, the field combination entering the no-go theorem inequality \eqref{combinogo} provides a good candidate. We now illustrate these points with several examples.\footnote{Given that the no-go theorems inequalities \eqref{combinogophi2} define field space directions, one may wonder whether those giving the same $c$ value involve the same direction. Looking at the no-gos that saturate the TCC bound, namely \ref{nogo2}, \ref{nogo5}, \ref{nogo7}, \ref{nogo9} and \ref{nogo10}, we can see that it is not always true. It only happens for the no-go \ref{nogo2} with $p=4$ and \ref{nogo10}, which involve the same linear combination of $V$ and its derivative. Overall, there is therefore not a single preferred scalar field direction.}\\

We start with examples where a single field enters the no-go theorem inequality \eqref{combinogo}. We begin with the no-go theorem \ref{nogo10}. Its inequality \eqref{nogo10ineq}, when compared to \eqref{combinogophi2}, indicates the field direction to be $\varphi_2 = \hat{\rho}$. Then, the large field distance is $\hat{\rho} \rightarrow \infty$, which also corresponds to $\rho \rightarrow \infty$. Here and in the following, we freeze the other fields at fixed values, taken to be $1$ for simplicity. In the framework of this no-go, the potential along the direction $\varphi_2=\hat\rho$ is then given, in Planckian units, by
\beq
V(\rho,\tau=1)=  - \rho^{-1} \frac{1}{2} {\cal R}_6 +\frac{1}{4} \rho^{-3} |H|^2 \approx - \rho^{-1} \frac{1}{2} {\cal R}_6 = \left(- \frac{1}{2} {\cal R}_6\right)\ e^{-\sqrt{\frac{2}{3}}\ \hat{\rho}} \ ,
\eeq
where the $\approx$ sign corresponds to taking the large field limit. The result is in perfect agreement with \eqref{Vcheck}. Another single field example is the no-go theorem \ref{nogo1}, for which we take $\varphi_2 = \hat{\tau}$. With a similar calculation, we obtain the potential in the large field limit $\tau \rightarrow \infty$
\beq
V(\rho=1,\tau) \approx \tau^{-2}  \frac{1}{2} \left( -  {\cal R}_6 +\frac{1}{2} |H|^2 \right)  = \left( -  {\cal R}_6 +\frac{1}{2} |H|^2 \right)\ e^{-\sqrt{2}\ \hat{\tau}} \ ,
\eeq
and verify again \eqref{Vcheck}. In both cases, the map is then verified.

We turn to an example with two fields in the no-go inequality \eqref{combinogo}. In that case, one has to introduce the appropriate field $\varphi_2$ and the orthogonal direction $\varphi_{2\bot}$. Given the inequalities \eqref{combinogophi2}, we propose to define these fields through the following rotation
\beq
\varphi_2 = \frac{b_1}{\sqrt{b_1^2 + b_2^2}}\, \hat{\varphi}^1 +  \frac{b_2}{\sqrt{b_1^2 + b_2^2}}\, \hat{\varphi}^2 \ , \qquad \varphi_{2\bot} = -\frac{b_2}{\sqrt{b_1^2 + b_2^2}}\, \hat{\varphi}^1 +  \frac{b_1}{\sqrt{b_1^2 + b_2^2}}\, \hat{\varphi}^2 \ . \label{varphi2}
\eeq
Indeed, this definition leads to the following field derivatives
\beq
b_1 \del_{\hat{\varphi}^1} + b_2 \del_{\hat{\varphi}^2} = \sqrt{b_1^2 + b_2^2} \, \del_{\varphi_2} \ ,
\eeq
that allows the rewriting \eqref{combinogophi2}, using also the formula for $c$ in \eqref{cformula}, while preserving the canonical normalization of the fields
\beq
(\del \varphi_2)^2 +  (\del \varphi_{2\bot})^2 = (\del \hat{\varphi}^1)^2 +  (\del \hat{\varphi}^2)^2 \ .
\eeq
We apply the definition \eqref{varphi2} to the no-go \ref{nogo9}, and obtain
\beq
\varphi_2 = \frac{1}{\sqrt{3}} \hat{\tau} + \sqrt{\frac{2}{3}} \hat{r} = \sqrt{\frac{2}{3}} \ln (\tau\, r) \ , \qquad \varphi_{2\bot} = - \sqrt{\frac{2}{3}} \hat{\tau} + \frac{1}{\sqrt{3}} \hat{r} = \frac{1}{\sqrt{3}} \ln(\tau^{-2}\, r) \ .
\eeq
Introducing
\beq
x= \tau\, r = e^{\sqrt{\frac{3}{2}}\, \varphi_2} \ , \quad y = \tau^{-2}\, r = e^{\sqrt{3}\, \varphi_{2\bot} } \ ,
\eeq
we rewrite the potential $V(r,\tau)$ given in \eqref{potredrt} in terms of $x,y$. We then set $y=1$, and send $x \rightarrow \infty$, corresponding to the large distance ($\varphi_2$) limit. We are then left with
\bea
V(r,\tau) \approx &\, \frac{1}{8} \delta^{ik} \,\delta^{jl} \, {f^{1}{}_{ij}} \, {f^{1}{}_{kl}}\, e^{\sqrt{\frac{2}{3}}\, \varphi_2} \\
+ &\, \frac{1}{2} \left(  -  {\cal R}_6 +  {\cal R}_{11} -\frac{1}{2}  \delta^{ik} \,\delta^{jl} \, {f^{1}{}_{ij}} \, {f^{1}{}_{kl}} + \frac{1}{2} |H^{(0)}|^2  - g_s \delta_1^{||} \frac{T_{10}}{p+1}  + \frac{1}{2} g_s^2  \sum_{q=0}^{5}|F_q^{(0)}|^2  \right) e^{-\sqrt{\frac{2}{3}}\, \varphi_2} \ ,\nn
\eea
where we kept the two dominant terms. The first term diverges, but it vanishes under the assumption of the no-go theorem \ref{nogo9}, namely that $f^{1}{}_{ij}=0$. Taking this into account, we obtain
\beq
V(r,\tau) \approx \frac{1}{2} \left(  -  {\cal R}_6 +  {\cal R}_{11} + \frac{1}{2} |H^{(0)}|^2  - g_s \delta_1^{||} \frac{T_{10}}{p+1}  + \frac{1}{2} g_s^2  \sum_{q=0}^{5}|F_q^{(0)}|^2  \right) e^{-\sqrt{\frac{2}{3}}\, \varphi_2} \ ,
\eeq
which verifies once again \eqref{Vcheck} and the map.

At this stage, one may wonder in general why the exponential behaviour of the potential \eqref{Vcheck} and thus the proposed map are actually verified in the examples considered. We give here a possible explanation. The potentials considered are sums of terms where the dependence on the fields $\varphi$ are simple powers. It turns out that the canonical fields are always given by $\hat{\varphi} \propto \ln \varphi$, such that the potential can be rewritten as a sum of terms with exponential dependence on the canonical fields. If we define the direction $\varphi_2$ of interest as in \eqref{combinogophi2}, then this is given by a linear combination of canonical fields, and thus appears as well in exponentials. If we freeze the other fields, we are left with the following
\beq
V(\varphi_2)= \sum_i a_i\, e^{- \gamma_i\, \varphi_2} \ ,\quad c\, V + \del_{\varphi_2} V =  \sum_i a_i\,  e^{- \gamma_i\, \varphi_2}\, (c - \gamma_i) \leq 0 \ ,
\eeq
where we define the $a_i$ such that all $\gamma_i$ are different. Usually, the appropriate combinations of the potential and its derivatives entering the no-go theorem are built in such a way that some specific contribution cancels-out. This implies that there exists one $\gamma_i$, say $i=1$, such that $c= \gamma_1$. In addition, the remaining terms in the combination are such that they all have the same definite sign (thanks to the assumptions of the no-go). To realise this, a possibility is that $a_{i} \geq 0$, which implies that $\gamma_{i} > c$, for $i\geq 2$. The opposite signs would also work, but they imply that $V$ would be negative in the large field limit, so we neglect for now this other option. With $\gamma_{i} > c$, for $i\geq 2$, and $c= \gamma_1$, we conclude that the potential has the appropriate form to match \eqref{Vcheck} in the large field distance, and the map is then verified. This scenario applies in the examples above, since the terms entering the constant $V_i$ are precisely those not appearing in the combination of $V$ and $\del V$. We hope to study further these questions in future work, as well as to analyse more examples, in order to get a better understanding of the behaviour of the potential in the large field limit.\\

Few more comments are now in order. What remains surprising about the map \eqref{ourconj} is that the two setups and the two fields are very different. We take for instance the setup${}_1$ to be a Minkowski compactification on a CY${}_3$ with an index $n$, indicating a specific asymptotic structure \cite{Grimm:2018ohb}; the field $\varphi_1$ in that case is a complex structure modulus. The setup${}_2$ is a compactification on e.g.~a group manifold with parallel $O_p/D_p$ sources, that obeys the assumptions of the various no-go theorems considered, and the field $\varphi_2$ depends on the no-go. For example, for $n=1$, one gets $c=2\lambda = \sqrt{2}$. The compactification setup${}_2$ can be specified to have $T_{10} \leq 0$, ${\cal R}_6 < 0$, and the map is then verified with $\varphi_2$ being the 4d dilaton, as shown above. Similarly, for $n=3$, one has $c=2\lambda = \sqrt{\frac{2}{3}}$. The setup${}_2$ can be either heterotic string at order $(\alpha')^0$ with e.g.~${\cal R}_6 < 0$ and $\varphi_2$ being the volume, or type II on a group manifold with $p\geq 4$, obeying assumptions of no-go \ref{nogo9}, with $\varphi_2$ a combination of a radius and the dilaton. In all these examples, the setups and the field directions are both very different.

This is related to the question on the physical reason underlying the existence of the proposed map. A nice interpretation would be a duality between string theory in different compactifications, in some specific regimes corresponding to some large field distances on both sides. For instance, the map between a CY framework and a complex structure modulus to a curved manifold and a K\"ahler modulus (radius, volume) is reminiscent of a T-duality or a (generalized) mirror symmetry. However such dualities typically require some restrictions on the compactification setups, and it is here unclear what is the extent or the generality of the above map. Candidate counter-examples to the de Sitter conjecture, e.g.~\cite{Danielsson:2011au, Andriot:2020wpp}, or to the strong anti-de Sitter distance conjecture, \cite{DeWolfe:2005uu, Junghans:2020acz, Marchesano:2020qvg}, may also indicate possible compactification setups that circumvent the applicability of such a map. In addition, having a relation between the two different quantities that are the mass and the potential remains unmotivated for now. Related ideas discussed in \cite{Ooguri:2018wrx, Grimm:2019ixq} could provide further hints. We would like to come back to these questions in future work, and hope that the present formulation of these ideas will still be helpful.

\vspace{0.4in}

\subsection*{Acknowledgements}

We would like to thank R.~Blumenhagen, T.~Grimm, J.~Knapp, T.~Schimannek, S.~Sethi and T.~Wrase warmly for useful exchanges during the completion of this project. D.~A.~acknowledges support from the Austrian Science Fund (FWF): project number M2247-N27. The work of N.~C.~is supported by an FWF grant with the number P 30265. D.~E.'s~work is supported by the FWF grant P30904-N27.

\newpage

\begin{appendix}

\section{4d four-forms and the scalar potential}\label{ap:4formgeneral}

\subsection{In $V(\rho,\tau,\sigma)$}\label{ap:4form}

When performing the dimensional reduction from the 10d type II supergravities action \eqref{S10d} to the final 4d one \eqref{S4d} with a scalar potential depending on $(\rho,\tau,\sigma)$, there is a non-trivial step involving 4d four-forms that we now detail. Because of 4d maximal symmetry, most fluxes can only be internal and can then be collected into a scalar potential $U$. Only $F_{4,5}^{10}$, depending on the theory, can have 4d components, which have to be proportional to ${\rm vol}_4$. Following \cite{Andriot:2016xvq, Andriot:2018ept}, we separate their 4d and 6d components as $F_4^{10} = F_4^4 + F_4$, $F_5^{10} = F_5^4 + F_5$. We further introduce $F_5^4=H_4 \w f_5$: $F_4^4$ and $H_4$ are 4d four-forms, $F_4$, $F_5$ and $f_5$ are purely 6d forms, $f_5$ being a one-form. Considering the fluctuations and dimensional reduction presented in section \ref{sec:framework}, and proceeding as in \cite{Andriot:2018ept}, we arrive in 4d Einstein frame at the following action
\bea
{\cal S}=  \int \d x^{4} \sqrt{|g_{4}|} \Bigg( & \frac{M_4^2}{2} {\cal R}_{4}  - {\rm kin} -  U(\rho,\sigma,\tau) \label{action4dap} \\
& \!\!\!\!\! - \frac{1}{2\kappa_{10}^2} \int \d^6 y \sqrt{|g_6^0|}\ g_s^{-2} \times \frac{1}{2} \tau^4 \rho^{3} g_s^2 \Bigg( |F_4^4|^{2} + \frac{1}{2}|H_4|^{2}\, \rho^{-1} \sum_{n=0,1} \sigma^{-An-B(1-n)} |f_5^{(n)}|^2 \Bigg) \Bigg) \nn
\eea
where for the conventions on the Planck mass and the kinetic terms we refer to section \ref{sec:framework}. In particular, the square of 4d four-form fluxes are in Einstein frame and we defined
\beq
U(\rho,\sigma,\tau) = \frac{1}{2\kappa_{10}^2} \int \d^6 y \sqrt{|g_6^0|}\ g_s^{-2} ( - \tau^{-2} \rho^{-1} {\cal R}_6(\sigma) + \dots ) \ ,
\eeq
where dots collect the purely internal fluxes and the sources contributions. One is tempted to identify $U$ together with the second line in \eqref{action4dap} as the complete 4d scalar potential $V$. This is in principle correct, but one has first to use the on-shell values (in a sense defined below) for the fluxes containing the 4d four-forms. Such a step requires particular attention. To understand the problem related to the on-shell evaluation of 4d four-forms, it will be enough to consider a simpler, purely 4d, action of the type
\beq
{\cal S}'= \int \d^4 x \sqrt{|g_{4}|} \left( {\cal R}_{4}  - \varphi |F_4|^2  \right) \ ,
\eeq
where $F_4 = \d C_3 $, $C_3$ being an abelian three-form gauge potential, and $\varphi$ a function of 4d coordinates (playing the role of our scalar fields). We recall that $\d^4 x \sqrt{|g_{4}|} |F_4|^2 = F_4 \w *_4 F_4 $. To obtain a background, as intended in the main text, one has to derive and solve the equations of motion. The equation of motion of $C_3$ is given by
\beq
\label{app:eomF4}
\d ( \varphi *_4 F_4 ) = 0 \ ,
\eeq
and its derivation requires a standard integration by parts that introduces a boundary term $2 \d (C_3 \w \varphi *_4 F_4)$. In particular, to get the equation of motion, this boundary term should vanish, implying $C_3 = 0$ on the boundary. While this requirement is usually supposed to be physical (fields vanish far away), it is formally not valid here, as $C_3 = 0$ is not a gauge invariant boundary condition. One should therefore consider an action corrected by this boundary term
\beq
{\cal S}''= \int \d^4 x \sqrt{|g_{4}|} \left( {\cal R}_{4}  - \varphi |F_4|^2  \right) + \int 2 \d (C_3 \w \varphi *_4 F_4)  \ .
\eeq
When taking the variation of $C_3$ no gauge-invariant boundary condition is now needed to get the equations of motion, since the problematic boundary term now cancels against the correction introduced. This correction should actually be introduced for any gauge theory, but usually the difference is not manifest, since the equations of motion are defined up to boundary terms. However, the specific case of 4d four-forms which are field strengths of gauge three-forms is an exception in this sense. Indeed, the equation \eqref{app:eomF4} indicates that $\varphi *_4 F_4$ is constant, therefore $\varphi F_4$ is not propagating any degree of freedom and should be viewed as an auxiliary field. It should then be integrated out and its on-shell value should be inserted in the action. This is precisely where having another term in the action makes a difference. On-shell, i.e.~using the equation of motion, the boundary term reduces to $\d (C_3 \w \varphi *_4 F_4) = F_4 \w \varphi *_4 F_4 = \d^4 x \sqrt{|g_{4}|} \varphi |F_4|^2 $, and the action becomes
\beq
{\cal S}''|_{{\rm on-shell}}= \int \d^4 x \sqrt{|g_{4}|} \left( {\cal R}_{4}  + \varphi |F_4|^2  \right)   \ ,
\eeq
i.e.~the $|F_4|^2$ term has flipped its sign. However, one should still evaluate this term on-shell. The equation of motion gives $\varphi *_4 F_4 = -f$, for some constant $f$ and with a minus sign for later convenience. Due to the 4d signature, one has $*_4^2 = -1$, so $F_4 = f \varphi^{-1} {\rm vol}_4$. Again because of the signature, the contraction denoted $|F_4|^2$ is actually negative, i.e.~here $|F_4|^2= - f^2 \varphi^{-2}$. We finally obtain
\beq
{\cal S}''|_{{\rm on-shell}}= \int \d^4 x \sqrt{|g_{4}|} \left( {\cal R}_{4}  -  \varphi^{-1} f^2 \right)   \ ,\ {\rm where}\ F_4 = f \varphi^{-1} {\rm vol}_4 \ .
\eeq
These results are well-known; some recent developments on the role of gauge three-forms in 4d, in connection to effective theories coming from string flux compactifications, can be found for example in \cite{Bielleman:2015ina, Carta:2016ynn, Farakos:2017jme, Herraez:2018vae, Lanza:2019xxg}.

Coming back to our initial action \eqref{action4dap}, we need to correct it in the same manner by the appropriate boundary term, and evaluate it on-shell. The function $\varphi$ corresponds to our set of scalar fields. To determine $f$ for our four-forms, we can further consider background values where for us $\varphi=1$. We proceed as in \cite{Andriot:2016xvq, Andriot:2018ept}, using the fact that the 4d space-time is maximally symmetric. In IIA we introduce a six-form $F_6^0$ so that the function $*_6 F_6^0$ captures all 6d dependence in the background: $F_4^{4\, 0}= {\rm vol}_4 \w *_6 F_6^0$. We insist that this $F_6^0$ is only a convenient notation and there is no such degree of freedom or flux in IIA; we are not using a democratic formalism here. In IIB, we can without loss of generality consider that all the 6d dependence has been captured by $f_5$, i.e.~$H_4^{0}= {\rm vol}_4$. Because of the background self-duality constraint of $F_5^{10}$, one must have $f_5^0=-*_6 F_5^0$. We then calculate in the aforementioned procedure the various integration constants $f^2$ and conclude that we should proceed with the following action
\bea
{\cal S}=  \int \d x^{4} \sqrt{|g_{4}|} \Bigg( & \frac{M_4^2}{2} {\cal R}_{4}  - {\rm kin} -  U(\rho,\sigma,\tau) \label{action4dap2} \\
& \!\!\!\!\!\!\!\!\!\!\!\!\!\!\! - \frac{1}{2\kappa_{10}^2} \int \d^6 y \sqrt{|g_6^0|}\ g_s^{-2} \times \frac{1}{2} g_s^2 \tau^{-4} \Bigg(  \rho^{-3} |F_6^0|^{2} + \frac{1}{2} \rho^{-2} \sum_{n=0,1} \sigma^{An+B(1-n)} |(*_6 F_5^0)^{(n)}|^2 \Bigg) \Bigg) \nn
\eea
In other words, the full scalar potential $V$ to be considered is
\bea
& {\cal S}=  \int \d x^{4} \sqrt{|g_{4}|} \Bigg( \frac{M_4^2}{2} {\cal R}_{4}  - {\rm kin} -  V(\rho,\sigma,\tau)\Bigg) \label{action4dap3} \\
& V = U  + \frac{1}{2\kappa_{10}^2} \int \d^6 y \sqrt{|g_6^0|}\ g_s^{-2} \times \frac{1}{2} g_s^2 \tau^{-4} \left(  \rho^{-3} |F_6^0|^{2} + \frac{1}{2} \rho^{-2} \sum_{n=0,1} \sigma^{An+B(1-n)} |(*_6 F_5^0)^{(n)}|^2 \right) \ . \nn
\eea
From this one derives the 4d Einstein equation and traces it to obtain, in the background,
\beq
{\cal R}_4 = \frac{4}{M_4^2} V|_0 \ .\label{R4value}
\eeq

Let us now compare to \cite{Andriot:2018ept, Andriot:2019wrs}, where the scalar potential for $\sigma$ was first fully derived. Keeping in mind the different $M_4$ convention, the value of ${\cal R}_4$ \eqref{R4value} is precisely the one used there, matching the one obtained from 10d. The scalar potential $V$ used in \cite{Andriot:2018ept, Andriot:2019wrs}, however, presents differences in the $*_6 F_5$ and $F_6$ terms, since these terms were not evaluated on-shell as done above. Integrating them out following the above procedure changes the sign of the $*_6 F_5$ and $F_6$ terms, and inverts the powers of the scalars. As a consequence, the first derivatives, in the background, are not affected: for instance,
\beq
\del_{\tau} \left( - \tau^4 |F_6^0|^{2} \right)|_{\tau=1} = \del_{\tau} \left( \tau^{-4} |F_6^0|^{2} \right)|_{\tau=1} \ .
\eeq
This explains why all combinations of ${\cal R}_4$ and first derivatives in the background in \cite{Andriot:2018ept, Andriot:2019wrs} are still correct, and match the 10d equations of motion, as checked. The second derivatives however would be changed in these two terms, but those were barely playing any crucial role in those papers.

To simplify further the potential \eqref{action4dap3} in type IIB, one notes an equality, for each source size $p$, between some $\sigma^{-An-B(5-n)} |F_5^{0\, (n)}|^2$ and a $\sigma^{An'+B(1-n')} |(*_6 F_5^0)^{(n')}|^2$. Indeed, one has
\bea
& p=3:\ \sigma^0  |F_5^{0\, (n=0)}|^2 = \sigma^0 |(*_6 F_5^0)^{(n=0)}|^2 \\
& p=5:\ \sigma^{-4}  |F_5^{0\, (n=1)}|^2 = \sigma^{-4} |(*_6 F_5^0)^{(n=1)}|^2 \ ,\ \sigma^2 |F_5^{0\, (n=2)}|^2 = \sigma^2 |(*_6 F_5^0)^{(n=0)}|^2\nn\\
& p=7:\ \sigma^{-2}  |F_5^{0\, (n=3)}|^2 = \sigma^{-2} |(*_6 F_5^0)^{(n=1)}|^2 \ ,\ \sigma^4 |F_5^{0\, (n=4)}|^2 = \sigma^4 |(*_6 F_5^0)^{(n=0)}|^2\nn\\
& p=9:\ \sigma^0  |F_5^{0\, (n=5)}|^2 = \sigma^0 |(*_6 F_5^0)^{(n=1)}|^2 \nn
\eea
This implies that $\sum_n \sigma^{-An-B(5-n)} |F_5^{0\, (n)}|^2 = \sum_{n} \sigma^{An+B(1-n)} |(*_6 F_5^0)^{(n)}|^2$, thus simplifying these terms. Another way to see this simplification is to note that $-An_{F_5}-B(5-n_{F_5}) = An_{*_6F_5}+B(1-n_{*_6F_5})$, or equivalently to $n_{F_5} = p-3 - n_{*_6F_5}$. With this, we finally obtain the potential
\bea
V = \frac{1}{2\kappa_{10}^2} g_s^{-2} & \int \d^6 y \sqrt{|g_6^0|} \bigg[ - \tau^{-2} \bigg( \rho^{-1} {\cal R}_6(\sigma) -\frac{1}{2} \rho^{-3} \sum_n \sigma^{-An-B(3-n)} |H^{(n)0}|^2 \bigg) \\
& \phantom{ \int \d^6 y \sqrt{|g_6^0|} \bigg[ } - g_s \tau^{-3} \rho^{\frac{p-6}{2}} \sigma^{B\frac{p-9}{2}} \frac{T_{10}^0}{p+1} \nn\\
& \phantom{ \int \d^6 y \sqrt{|g_6^0|} \bigg[ } +\frac{1}{2} g_s^2 \tau^{-4} \bigg(  \sum_{q=0}^{5} \rho^{3-q} \sum_n  \sigma^{-An-B(q-n)} |F_q^{(n)0}|^2  +  \rho^{-3} |F_6^0|^2 \bigg) \bigg] \ , \nn
\eea
which corresponds to \eqref{pot0N} in the main text. Note that the $F_6$ term could be incorporated in the last sum by verifying that for each source size $p$, the corresponding $\sigma$ power is vanishing. Indeed, one has for $q=6$ that $-An-B(q-n) = 6 (n - (p-3))$ and, for a six-form, there is only one $n$ which equates the number of internal source directions, namely $p-3$. For clarity of the $\sigma$ dependence however, we leave the $F_6$ term aside.

\subsection{In $V(r,\tau)$}\label{ap:4formr}

We proceed similarly for the fluctuations considered in section \ref{sec:radius}, namely the radius $r$ and the 4d dilaton $\tau$. Starting from the 10d action, we reach as above the following 4d action
\bea
{\cal S}=  \int \d x^{4} \sqrt{|g_{4}|} \Bigg( & \frac{M_4^2}{2} {\cal R}_{4}  - {\rm kin} -  U(r,\tau) \label{action4drap} \\
& \!\!\!\!\! - \frac{1}{2\kappa_{10}^2} \int \d^6 y \sqrt{|g_6^0|}\ g_s^{-2} \times \frac{1}{2} \tau^4 r g_s^2 \left( |F_4^4|^{2} + \frac{1}{2}|H_4|^{2}\, ( |f_5^{(0)}|^2 + r^{-2} |f_5^{(1)}|^2 ) \right) \Bigg) \nn
\eea
where the square of 4d four-form fluxes are in Einstein frame and we defined
\beq
U(r,\tau) = \frac{1}{2\kappa_{10}^2} \int \d^6 y \sqrt{|g_6^0|}\ g_s^{-2} \left( - \tau^{-2} {\cal R}_6(r) + \frac{g_s^{2}}{2} \tau^{-4} r \, \frac{1}{2} |F_5|^2(r)  + \dots \right) \ ,
\eeq
with dots standing for the rest of the contributions. It is straightforward to proceed as explained above: one must add the total derivative term, and integrate-out the four-form fluxes, by evaluating the resulting action on-shell. This operation results in the following changes in the previous action
\bea
& \tau^4 r |F_4^4|^{2} \rightarrow -  \tau^{-4} r^{-1} |F_4^4|^{2} = + \tau^{-4} r^{-1} |F_6|^{2} \ ,\\
& \tau^4 r |H_4|^{2}\, ( |f_5^{(0)}|^2 + r^{-2} |f_5^{(1)}|^2 ) \rightarrow - \tau^{-4}  |H_4|^{2}\, ( r^{-1} |f_5^{(0)}|^2 + r |f_5^{(1)}|^2 ) \nn\\
& \phantom{\tau^4 r |H_4|^{2}\, ( |f_5^{(0)}|^2 + r^{-2} |f_5^{(1)}|^2 ) }  = + \tau^{-4} \, ( r^{-1} |(*_6 F_5)^{(0)}|^2 + r |(*_6 F_5)^{(1)}|^2 ) \ ,\nn
\eea
with fluxes now on-shell, and we dropped the superscript ${}^0$. As above, we further simplify the $F_5$ terms by noticing that
\beq
|(*_6 F_5)^{(0)}|^2 = |F_5^{(1)}|^2 \ ,\  |(*_6 F_5)^{(1)}|^2 = |F_5^{(0)}|^2 \ .
\eeq
This eventually leads us to the following potential
\bea
V(r,\tau) & =U(r,\tau) +\frac{1}{2\kappa_{10}^2} \int \d^6 y \sqrt{|g_6^0|}\ g_s^{-2} \times \frac{g_s^2}{2}  \tau^{-4} \left(   r^{-1} |F_6|^{2} +  \frac{1}{2} \, ( r^{-1} |(*_6 F_5)^{(0)}|^2 + r |(*_6 F_5)^{(1)}|^2 ) \right) \nn \\
& = \frac{1}{2\kappa_{10}^2} \int \d^6 y \sqrt{|g_6^0|}\ g_s^{-2} \left( - \tau^{-2} {\cal R}_6(r) + \frac{g_s^{2}}{2} \tau^{-4} \left( r |F_5^{(0)}|^2+ r^{-1} |F_5^{(1)}|^2 + r^{-1} |F_6|^{2} \right)  + \dots \right) \ ,\nn
\eea
fully expressed in \eqref{potrtau0}. As before, the $F_5$ terms simplify and get incorporated to the sum of other fluxes. The $F_6$ could be included as well, since a six-form is necessarily such that $F_6^{(0)}=0$.

Given this simple final result, one may question the need to go through this treatment of 4d four-forms, and not simply fluctuate $F_5$ and $F_6$ from the start, as done e.g.~in type IIA in \cite{Hertzberg:2007wc}. For $F_6$ we are reluctant to do so because there is fundamentally no six-form in type IIA 10d theory, as well as no square of a six-form which would involve six inverse internal metrics; here it is only introduced as a convenient notation through $*_6 F_6$ to capture an internal function. Similarly for $F_5$, the 10d starting action is different than simply $|F_5|^2$. In addition, $F_5^{10}$ is also required to satisfy the self-duality constraint, which we prefer to impose on an on-shell background flux value, that forces us to proceed as we did. The final potential remains the same as in \cite{Hertzberg:2007wc} for IIA, and as in follow-up papers.

\end{appendix}

\newpage

\providecommand{\href}[2]{#2}\begingroup\raggedright
\endgroup

\end{document}